\documentclass[12pt]{article}
\usepackage{epsf}
\usepackage{graphics}
\usepackage[dvips]{graphicx}
\usepackage{epsfig}
\usepackage{amssymb}
\usepackage{caption}
\newcommand{\lsc}{\Lambda _\chi}
\newcommand{\cpt}{$\chi$PT }
\newcommand{\sss}{\scriptscriptstyle }
\newcommand{\ga}{g_{\sss A}}

\newcommand{\vsigone}{{\vec\sigma^{\sss 1}}}
\newcommand{\vsigtwo}{{\vec\sigma^{\sss 2}}}
\newcommand{\veps}{{\vec\epsilon}}
\newcommand{\vepsprime}{{\vec\epsilon\, '}}
\newcommand{\vkay}{{\vec k}}
\newcommand{\vkayprime}{{{\vec k}^{\, \prime}}}
\newcommand{\vpee}{{\vec p}}
\newcommand{\vpeeprime}{{{\vec p}\, '}}
\newcommand{\vsigma}{{\vec\sigma}}

\newcommand{\lsim}{\, \, \raisebox{-0.8ex}{$\stackrel{\textstyle <}{\sim}$ }}

\newcommand{\qtwoline}{(\vec{p}-\vpeeprime + \smfrac{1}{2}\vec{q}^{\,})}
\newcommand{\qoneline}{(\vec{p}-\vpeeprime - \smfrac{1}{2}\vec{q}^{\,})}
\newcommand{\qnonline}{(\vec{p}-\vpeeprime + \vec{k}_+)}

\def\Z{{\cal Z}}
\def\mn{M}
\def\mpi{m_\pi}
\def\fpi{f_\pi}
\def\half{{\textstyle {1\over 2}}}
\def\d{{\rm d}}
\def\t{{\bf t}}
\def\ombar{{\overline\omega}}

\def\sigbol{\vec{\sigma}}

\def\qvec{\vec{q}}
\def\pbar{\vec{p}_+}
\def\pq{\vec{p}_+ \cdot \vec{k}_+}
\def\beq{\begin{equation}}
\def\eeq{\end{equation}}
\def\smfrac#1#2{{\textstyle {#1\over #2}}}

\begin{document}
\newpage
\baselineskip 16pt plus 2pt minus 2pt

\thispagestyle{empty}

\par
\topmargin=-1cm      



\vspace{0.6cm}
\begin{centering}
{\Large\bf Compton scattering on the proton, neutron, and
deuteron in chiral perturbation theory to $O(Q^4)$}

\vspace{1.0cm}

{{\bf S.R.~Beane}$^{1,2}$,
{\bf M.~Malheiro}$^{3}$, {\bf J.A.~McGovern}$^{4}$,\\
{\bf D.R.~Phillips}$^{5}$, and
{\bf U.~van Kolck}$^{6,7,8}$ }\\
\vspace{0.8cm}
{\sl $^{1}$ Physics Department, University of New Hampshire,\\
209E DeMeritt Hall, 9 Library Way, Durham, NH 03824, USA}\\
\vspace{5.0pt}
{\sl $^{2}$ Jefferson Laboratory,
Newport News, VA 23606, USA}\\
\vspace{5.0pt}
{\sl $^{3}$Instituto de F\'{\i}sica, Universidade Federal Fluminense,\\
24210-340, Niter\'oi, R.J., Brazil } \\
\vspace{5.0pt}
{\sl $^{4}$Department of Physics and Astronomy, University of Manchester,\\
Manchester M13 9PL, UK} \\
\vspace{5.0pt}
{\sl $^{5}$Department of Physics and Astronomy, Ohio University,\\
Athens, OH 45701, USA}\\
\vspace{5.0pt}
{\sl $^{6}$Department of Physics, University of Arizona,\\
Tucson, AZ 85721, USA}\\
\vspace{5.0pt}
{\sl $^{7}$RIKEN BNL Research Center, Brookhaven National Laboratory,} \\
{\sl Upton, NY 11973, USA}\\
\vspace{5.0pt}
{\sl $^{8}$Department of Physics, University of Washington,\\
Box 351560, Seattle, WA 98195, USA}\\
\end{centering}

\vspace{0.8cm}

\begin{center}
\begin{abstract}

\vspace*{0.1cm}

\noindent
We study Compton scattering in systems with A=1 and 2 using chiral
perturbation theory up to fourth order. For the proton we fit the two
undetermined parameters in the $O(Q^4)$ $\gamma$p amplitude of
McGovern to experimental data in the region $\omega,\sqrt{|t|} \leq
180$ MeV, obtaining a $\chi^2/{\rm d.o.f.}$ of 133/113.  This yields a
model-independent extraction of proton polarizabilities based solely
on low-energy data:
$\alpha_p=(12.1 \pm 1.1~({\rm stat.}))_{-0.5}^{+0.5}~({\rm theory})$ and $\beta_p=(3.4 \pm 1.1~({\rm stat.}))_{-0.1}^{+0.1}~({\rm theory})$, both 
in units of $10^{-4}~{\rm fm}^3$.
We also compute Compton scattering on deuterium to $O(Q^4)$.  The
$\gamma$d amplitude is a sum of one- and two-nucleon mechanisms, and
contains two undetermined parameters, which are related to the
isoscalar nucleon polarizabilities. We fit data points from three
recent $\gamma$d scattering experiments with a $\chi^2/{\rm
d.o.f.}=26.6/20$, and find
$\alpha_N=(13.0 \pm 1.9~({\rm stat.}))_{-1.5}^{+3.9}~({\rm theory})$
and a $\beta_N$ that is consistent with zero within sizeable error bars.
\end{abstract}

\vspace*{15pt}
\noindent
PACS nos.: 13.60.Fz, 12.39.Fe, 25.20.-x, 12.39.Pn, 21.45.+v

\end{center}

\newpage

\section{Introduction}
\label{sec-intro}

Compton scattering on the proton, neutron, and deuteron provides a
window on the internal dynamics of these strongly-interacting
systems. As this is an electromagnetic process, the photon-target
interaction can be treated using well-controlled approximations.  At
the same time Compton scattering is governed by a different set of
target structure functions than, say, electron scattering, and so
probes complementary aspects of the strong nuclear force. For example,
in the case of the proton, quark models indicate that even at quite
low energies $\gamma$p scattering can reveal distinctive information
about the charge and current distributions produced by the nucleon's
quark substructure~\cite{holstein,CK92}.

The deuteron represents a different theoretical challenge. Its
structure is governed by the nucleon-nucleon ($NN$) interaction. A
number of models of this interaction exist which very accurately
describe experimental data on $NN$ scattering at laboratory energies
below 350 MeV. However, these successful approaches differ
significantly in their underlying assumptions. They range from models
which are almost purely phenomenological~\cite{argonne} to
multi-boson-exchange models \cite{Bonnrep,Nijm93,NijmESC} that make
detailed assumptions about how QCD plays out in the $NN$ system,
e.~g. the role of the pomeron~\cite{Nijm93,NijmESC}.  Clearly, $NN$
scattering alone cannot completely constrain the form of strong
interactions at low energies.  Deuteron Compton scattering is a way of
probing the deuteron bound-state dynamics generated by different
models.  Furthermore, any theoretical treatment of this reaction must
incorporate not just photon interactions with the target via the
one-body Compton process but two-body mechanisms as well: the input to
a calculation of the $\gamma$d amplitude includes the $\gamma N$
amplitude plus Compton-scattering two-body currents. These two-body
currents should be constructed in a way that is consistent with both
the single-nucleon Compton dynamics and the $NN$ potential used to
generate the nuclear bound state.

Over the last several years a framework~\cite{bkreview} based on
effective field theory (EFT)~\cite{kreview} has been developed that
allows such consistency. Importantly, EFTs also satisfy the symmetry
constraints of QCD while making only minimal theoretical assumptions. This
framework can be applied to the Compton processes we are interested in,
provided that the photon energy is well below the chiral symmetry
breaking scale, $\lsc\sim 4 \pi f_\pi \sim M \sim {m_\rho}$. In this
regime photon-nucleon and photon-deuteron scattering are governed by
the approximate chiral $SU(2)_L \times SU(2)_R$ symmetry of QCD. The
spontaneous breaking of this symmetry to isospin $SU(2)_V$ results in
the existence of three Goldstone bosons, which are identified with the
pions ($\pi^+$, $\pi^-$, and $\pi^0$).

An EFT can be built from the most general Lagrangian involving pions
and nucleons~\footnote{The simplest form of the EFT results if the
Delta isobar is integrated out of the theory, although this tends to
limit the range of energies over which data can be described.} which
is constrained only by approximate chiral symmetry and the relevant
space-time symmetries.  In this EFT, known as chiral perturbation theory
($\chi$PT), pions interact through vertices with a countable number of
derivatives and/or insertions of the quark mass matrix.  Therefore
$S$-matrix elements can be expressed as simultaneous expansions in
powers of momenta and pion masses over the characteristic scale of
physics that is not included explicitly in the EFT, $\Lambda_\chi$. A generic
amplitude can be written as:
\begin{equation}
T = C(\Lambda_\chi) \, \sum_\nu \, \left(\frac{Q}{\Lambda_\chi}\right) ^\nu
             {\cal F}_\nu (Q/m_\pi),
\label{pionfulT}
\end{equation}
where $Q$ represents the external momenta and/or the pion mass,
${\cal F}_\nu$ is a calculable function with ${\cal F}_\nu(1)$ 
of order one, $C$ is an overall normalization factor, and $\nu$ is a
counting index.

\cpt has proven remarkably successful, with a number of single-nucleon
processes \cite{bkmrev}---including $\gamma$p scattering 
\cite{ulf1,ulf2,judith}---now computed to several nontrivial
orders.  
In $\chi$PT, as in any EFT, the price to be
paid for the absence of model assumptions is a lack of detailed
information about physics at distance scales much shorter than the
wavelength of the probe. This price translates into the presence
in $\chi$PT amplitudes of certain undetermined
interaction strengths, or ``low-energy constants'' (LECs). 
These constants can be estimated using models of the
short-distance physics, but in the purest form of \cpt they should be
fitted to experimental data.

There has been intense recent effort dedicated to extending \cpt to
processes with more than a single nucleon \cite{bkreview}.  The
fundamental difficulty in making this extension is the treatment of
the effects of nuclear binding.  At momenta comparable to the pion
mass the few-nucleon EFT first proposed by Weinberg~\cite{weinnp}
still provides a paradigm for the application of EFT to nuclear
systems. Use of this few-nucleon EFT has developed to the point where
computations of processes involving two nucleons with any number of
pions and photons can be carried out at levels of precision comparable
to those achieved using \cpt in the single-nucleon
sector~\cite{hybrid}. The present paper is part of this ongoing
program~\cite{therestofus,allofus}.

During roughly the same period over which EFT has developed as a tool
for analyzing low-energy QCD dynamics, experimental facilities with
tagged photons have made possible a new generation of Compton
scattering experiments which probe the low-energy structure of
nucleons and nuclei. An extensive database now exists for Compton
scattering on the proton at photon energies below 200
MeV~\cite{pdata}, with this database being considerably enhanced 
by the recent data taken in the TAPS setup at MAMI by Olmos de
L\'eon {\it et al.}~\cite{mainz}. In the nuclear case the past decade
has seen a set of experiments that establish---for the first time---a
modern data set for Compton scattering on the deuteron.  Data now
exist for coherent $\gamma d\rightarrow \gamma d$ from 49 to 95 MeV
\cite{lucas,SAL,lund}, and for quasi-free $\gamma d\rightarrow \gamma
pn$ from 200 to 400 MeV \cite{neutpol3,kolb,kossert}.

These experimental and theoretical developments come together in
studies of Compton scattering on the $A=1$ and $A=2$ system in
$\chi$PT.  Nucleon Compton scattering has been studied in \cpt to
$O(Q^3)$~\cite{bkmrev,ulf1} and $O(Q^4)$~\cite{ulf2,judith}.  To
$O(Q^3)$, there are {\it no} undetermined parameters in the $\gamma$p
amplitude, so \cpt makes predictions for $\gamma$p scattering.  These
agree reasonably well with low-energy differential cross-section
data~\cite{bkmrev,babusci}.  At the next order in the chiral
expansion, $O(Q^4)$, there are four undetermined parameters, two for
$\gamma$p scattering and two for $\gamma$n scattering. These
counterterms account for short-range contributions to nucleon
structure. Minimally one needs four pieces of experimental data to
determine their values, and once they are fixed \cpt~makes improved,
model-independent predictions for Compton scattering on protons and
neutrons~\cite{judith}.  Below we use low-energy proton data in order
to obtain the two short-distance parameters associated with the
proton, and so make a completely model-independent determination of
the \cpt $\gamma$p amplitude up to fourth order in small quantities.

In the $A=2$ system, the amplitude for coherent Compton scattering on
the deuteron was computed to $O(Q^3)$ in Ref.~\cite{therestofus}.
There are {\it no} free parameters at this order. The corresponding
cross section is in good agreement with the Illinois data~\cite{lucas}
at 49 and 69 MeV, but underpredicts the SAL data~\cite{SAL} at 95
MeV. The calculation of Ref.~\cite{therestofus} also yields cross
sections which agree well with the more recent Lund data~\cite{lund}.
In the present paper we extend this calculation to $O (Q^4)$.  At this
order coherent $\gamma$d scattering is sensitive to two isoscalar
combinations of the four free parameters in the $O(Q^4)$ $\gamma N$
amplitude. We will show that fitting these two parameters to
experiment yields $\gamma$d cross sections which are in good agreement
with the existing data for momentum transfers and photon energies
below 160 MeV.  A summary of earlier $O(Q^4)$ \cpt results for both
$\gamma$p and $\gamma$d scattering was presented in
Ref.~\cite{allofus}. The current paper contains a considerably more
detailed discussion. Because of errors in the fits to $\gamma$d data
reported in Ref.~\cite{allofus} the central values found here for the
isoscalar polarizabilities differ significantly from those given in
Ref.~\cite{allofus} (see Section~\ref{sec-gdfits} for a full
explanation).

Very low-energy nucleon Compton scattering (with $\omega$
significantly below $m_\pi$) has been a focus of particular theoretical
interest.  At these energies, the \cpt $\gamma N$ amplitude can be
expanded in powers of $\omega/m_\pi$. This yields a $T$-matrix that is a simple
power series in $\omega$, with coefficients which are functions of \cpt
parameters. In the nucleon rest frame,
\begin{equation}
T=\vepsprime \cdot \veps
\left(-\frac{{\cal Z}^2 e^2}{M} + 4 \pi \alpha \omega \omega'\right)
+ 4 \pi \beta \, \vepsprime\times \vec{k}' \cdot \veps\times \vec{k} 
+ \ldots.
\label{pionlessT}
\end{equation}
Here the ellipses represent higher powers of energy and momentum as well as
relativistic corrections, and $\veps$ ($\vepsprime$) is the
polarization vector of the initial- (final-) state photon and $\vec{k}$
($\vkayprime$) is its three-momentum.  The first term in this series
is a consequence of gauge invariance, and is the Thomson limit for
Compton scattering on a target of mass $M$ and charge ${\cal Z}
|e|$. Meanwhile, the coefficients of the second and third terms are
the nucleon electric and magnetic polarizabilities, $\alpha$ and
$\beta$. To $O(Q^3)$ in \cpt they are completely given by
pion-loop effects~\cite{ulf1}:
\begin{eqnarray}
\alpha_p=\alpha_n=\frac{5 e^2 g_A^2}{384 \pi^2 f_\pi^2 m_\pi} &=&12.2 \times
10^{-4} \, {\rm fm}^3; \nonumber\\
\beta_p=\beta_n=\frac{e^2 g_A^2}{768 \pi^2 f_\pi^2 m_\pi}&=& 1.2 \times
10^{-4} \, {\rm fm}^3, \label{eq:betaOQ3}
\end{eqnarray}
with $g_A$ the axial coupling of the nucleon and
$f_\pi$ the pion decay constant.  We emphasize that
the polarizabilities are {\it predictions} of \cpt at this order: they
diverge in the chiral limit because they arise from pion-loop
effects. In less precise language, \cpt tells us that polarizabilities
are dominated by the dynamics of the long-range pion cloud
surrounding the nucleon, rather than by short-range dynamics, and thus
should provide a sensitive test of chiral dynamics.

At $O(Q^4)$ the four undetermined parameters alluded to above
contribute to the polarizabilities~\cite{ulf2}.  They can be
understood as representing short-distance ($r \ll 1/m_\pi$) effects in
$\alpha$ and $\beta$.  The results from our fits to proton Compton
data provide the first completely model-independent determination of
proton polarizabilities. These results, together with an analysis of
various remaining sources of theoretical uncertainty, are given in
Section~\ref{sec-protres} below.

While data from Compton scattering on hydrogen targets can be used to
extract the proton polarizabilities $\alpha_p$ and $\beta_p$, the
absence of dense, stable, free neutron targets requires that the
neutron polarizabilities $\alpha_n$ and $\beta_n$ be extracted from
scattering on deuterium (or some other nuclear target).  Coherent
Compton scattering on deuterium depends on the polarizability
combinations $\alpha_N \equiv (\alpha_p + \alpha_n)/2$ and $\beta_N
\equiv (\beta_p + \beta_n)/2$ through interference between the
polarizability pieces of the $\gamma N$ amplitude and the proton
Thomson term. Thus $\alpha_N$ and $\beta_N$ can be extracted from
nuclear data, but to do so requires a consistent theoretical framework
that cleanly separates single-nucleon properties from multi-nucleon
effects. In the long-wavelength limit pertinent to polarizabilities
EFT provides a model-independent way to do exactly
this~\cite{bkreview,kreview,bkmrev}, and thus we would argue that \cpt
(or at least some systematic EFT) is an essential tool in the quest to
obtain $\alpha_N$ and $\beta_N$ from deuteron Compton data.  Our \cpt
analysis of $\gamma$d data in Section~\ref{sec-results} facilitates a
model-independent extraction of $\alpha_N$ and $\beta_N$, as well as a
quantitative assessment of the uncertainty associated with the freedom
to choose different $NN$ potentials when generating the deuteron bound
state.

Our results for both $\gamma$d cross sections and extracted values of
$\alpha_N$ and $\beta_N$ agree qualitatively with potential-model
treatments of $\gamma$d scattering at these
energies~\cite{wilbois,levchuk,jerry}. The different potential-model
calculations all yield similar results if similar input is
supplied---in particular in the form of comprehensive meson-exchange
currents. Typically though, the one- and two-nucleon Compton
scattering mechanisms in potential models are not derived in a
mutually-consistent fashion. The calculations of
Refs.~\cite{wilbois,levchuk,jerry} also tend to incorporate
significantly more model assumptions about short-distance physics than
are necessary in an EFT approach.

We note that calculations of Compton scattering on the deuteron in
EFTs other than the one employed here exist in the literature.  For
example, in Ref.~\cite{griess}, an EFT where pions are integrated out
is employed.  This EFT is under good theoretical control and is a very
precise computational tool, but it is limited in its utility to
momenta well below the pion mass.  In Ref.~\cite{Ch98} a different power
counting~\cite{Ka98} from ours was used in which pion exchange is
treated in perturbation theory.  While this power counting is
well-adapted to processes at momenta below the pion mass, it is now
known to break down at quite low energies~\cite{FMS}.

The rest of the paper is structured as follows.  In
Section~\ref{sec-bcpt} we summarize those aspects of \cpt which are
directly relevant to our calculation, including the power-counting
scheme and the pertinent terms in the chiral Lagrangian.  In
Section~\ref{sec-prottheory} we explain how this theory is applied to
Compton scattering on the proton up to $O(Q^4)$, reviewing the results
of Ref.~\cite{judith}. In Section~\ref{sec-protres} we fit the two
unknown parameters in the $O(Q^4)$ \cpt $\gamma$p amplitude to
$\gamma$p scattering data with $\omega,\sqrt{|t|} \leq 180$ MeV.
In Section~\ref{sec-deuttheory} we turn to the A=2 system,
computing the Feynman amplitudes that contribute to photon scattering
on the two-nucleon system at $O(Q^4)$. In Section~\ref{sec-results} we
present results for differential cross sections for Compton scattering
on unpolarized deuteron targets.  We discuss higher-order effects and
the extent to which a determination of neutron polarizabilities from
elastic $\gamma$d is possible. Finally, in
Section~\ref{sec-conclusion} we summarize our work, and identify
future directions for the theoretical study of this reaction.

\section{Baryon Chiral Perturbation Theory}
\label{sec-bcpt}

In this section we briefly discuss the effective chiral Lagrangian
underlying our calculations and the corresponding power counting.

\subsection{Effective Lagrangian}

QCD has an approximate chiral $SU(2)_L \times SU(2)_R$ symmetry that
is spontaneously broken at a scale $\lsc\sim 1$ GeV.  Below this scale
there are no obvious small coupling constants in which to expand.
Effective field theory is the technique by which a hierarchy of scales
is developed into a perturbative expansion of physical observables.
Here we are interested in processes where the typical momenta of all
external particles is $p\ll\lsc$, so we identify our expansion
parameter as $p/\lsc$.

In a system with broken symmetries this technique is especially
powerful. When a continuous symmetry is spontaneously broken there are
always massless Goldstone modes which dominate the low-energy
dynamics.  If the chiral symmetry of QCD were exact, $p/\lsc$ would be
the only expansion parameter.  However, in QCD $SU(2)_L \times
SU(2)_R$ is softly broken by the small quark masses. This explicit
breaking implies that the pion has a small mass in the low-energy
theory.  Since ${m_\pi}/\lsc$ is then also a small parameter, we have
a dual expansion in $p/\lsc$ and ${m_\pi}/\lsc$.

We will limit ourselves to the region where $m_\pi \sim p < M_\Delta
-M$.  ($M$ and $M_\Delta$ stand for the nucleon and Delta-isobar mass,
respectively.)  In this regime we can keep only pions and nucleons as
active degrees of freedom, and the low-energy EFT is \cpt without an
explicit Delta-isobar field.  We take $Q$ to represent either a small
momentum {\it or} a pion mass, and seek an expansion in powers of
$Q/\lsc$.  We implicitly include the mass scale of the degrees of
freedom that have been integrated out---$M_\Delta -M$, $m_\rho$,
etc.--- in $\lsc$.  One might reasonably expect that the convergence
of the theory will improve upon the inclusion of an explicit Delta
field \cite{jenkinsetal,hemmert}.  Work on nucleon Compton scattering
along these lines is ongoing~\cite{PP03,Hi03}.

Although chiral symmetry is no longer a symmetry of the vacuum, it
remains a symmetry of the Lagrangian.  One can show \cite{CCWZ} that a
choice of pion fields is possible, in which all pion interactions are
either derivative or proportional to $m_\pi^2$. The symmetries of QCD
still permit an infinite number of such interactions though, and so it
is necessary to order the strong interactions according to the
so-called index of the interaction~\cite{weinnp}. For an interaction
labeled $i$ this is defined by:
\begin{equation}
{\Delta_i}\equiv {d_i}+{f_i}/2-2
\label{Deltai}
\end{equation}
with $d_i$ the sum of the number of derivatives or powers of $m_\pi$ and 
$f_i$ the number of nucleon fields.
Since pions couple to each other and to nucleons through either
derivative interactions or quark masses, there is a lower bound on the
index of the interaction: ${\Delta _i}\geq 0$. 
We write the effective Lagrangian as
\begin{equation}
{\cal L} =\sum_{\Delta=0}^{\infty} {\cal L}^{(\Delta)}.
\end{equation}

So far this only accounts for strong interactions. The electromagnetic
field can enter ${\cal L}$ either through minimal substitution or by
the addition of terms involving the electromagnetic field strength
tensor. The former simply has the effect of replacing a derivative by
a factor of the charge $e$. And in either case we now have another
expansion: one in the small electromagnetic coupling $\alpha_{\rm
em}=e^2/4\pi$. However, since we work only to leading order in this
expansion it is convenient to enlarge the definition of $d_i$ so that it
includes powers of $e$, and then continue to classify interactions
according to the index $\Delta _i$ defined by Eq. (\ref{Deltai}).

The technology that goes into building an effective
Lagrangian and extracting the Feynman rules is standard by now and
presented in great detail in Ref. \cite{bkmrev}.
Here we list only the elements necessary for our calculations.

The pion triplet is contained in a matrix field
\begin{equation}
\Sigma =\xi^2 \equiv \sqrt{1-\frac{\vec\pi^2}{f_\pi^2}} 
        +i \frac{{\vec\pi}\cdot{\vec\tau}}{f_\pi},
\end{equation}
where $f_\pi$ is the pion decay constant, for which 
we adopt the value $f_\pi=92.4$ MeV.  Under
$SU(2)_L\times SU(2)_R$, $\Sigma$ transforms as $\Sigma\rightarrow
L\Sigma{R^\dagger}$ and $\xi$ as $\xi\rightarrow
L\xi{U^\dagger}=U\xi{R^\dagger}$; here $L$($R$) is an element of
$SU(2)_L$ ($SU(2)_R$), and $U$ is defined implicitly.  It is
convenient to assign the nucleon doublet $N$ the transformation
property $N\rightarrow UN$.
With ${\cal Z}=(1+{\tau _3})/2$ we can write the pion covariant derivative as
\begin{equation}
{D_\mu}\Sigma={\partial _\mu}\Sigma -ie{{\cal A}_\mu}[{\cal Z},\Sigma].
\end{equation}
Out of $\xi$ one can construct
\begin{eqnarray}
{V_\mu}&=&\frac{1}{2}[{\xi ^\dagger}({\partial _\mu}-ie{{\cal A}_\mu}{\cal Z})\xi +
 \xi({\partial _\mu}-ie{{\cal A}_\mu}{\cal Z}){\xi ^\dagger}] \\
{A_\mu}&=&\frac{i}{2}[{\xi ^\dagger}({\partial _\mu}-ie{{\cal A}_\mu}{\cal Z})\xi -
 \xi({\partial _\mu}-ie{{\cal A}_\mu}{\cal Z}){\xi ^\dagger}], 
\end{eqnarray}
which transform as ${V_\mu}\rightarrow U{V_\mu}{U^\dagger} +
U{\partial _\mu}{U^\dagger}$ and ${A_\mu}\rightarrow
U{A_\mu}{U^\dagger}$ under $SU(2)_L\times SU(2)_R$.  $V_\mu$ is used
to build covariant derivatives of the nucleon,
\begin{equation}
D_\mu N= ({\partial _\mu}+{V_\mu})N.
\end{equation}
(Ref.~\cite{Fe00} uses $\Gamma_\mu$ where we have used $V_\mu$ and 
$\smfrac{1}{2} u_\mu$ for our $A_\mu$.)
The electromagnetic field enters not only through minimal coupling in
the covariant derivatives, but also through
\begin{equation}
f_{\mu\nu}= e({\xi ^\dagger}{\cal Z}\xi + \xi {\cal Z}{\xi ^\dagger}){F_{\mu\nu}},
\end{equation}  
where ${F_{\mu\nu}}=\partial_\mu A_\nu - \partial_\nu A_\mu$ is the
usual electromagnetic field strength tensor.

Because the nucleon mass is large, $M \gg Q$, it plays no dynamical
role: nucleons are non-relativistic objects in the processes we are
interested in. The field $N$ can be treated as a heavy field of
velocity $v$ in which on-mass-shell propagation through $\exp
(iM{v\cdot x})$ has been factored out.  In the rest frame of the
nucleon, the velocity vector ${v^\mu}=(1,\vec{0})$.  The spin operator
is denoted by $S^\mu$, and in the nucleon rest frame
$S^\mu=(1/2)(0,\vec{\sigma})$. 
This formulation of \cpt is sometimes called heavy-baryon chiral
perturbation theory (HB$\chi$PT). 

With these ingredients, it is straightforward to construct the
effective Lagrangian.
As will become clear later, in
order to calculate Compton scattering
to $O(Q^4)$ we need to consider interactions with $\Delta_i \le 3$.

The leading-order Lagrangian is
\begin{eqnarray}
{\cal L}^{(0)}& = &
     \frac{1}{4}{f_\pi^2}{\rm Tr}({D_\mu}{\Sigma ^\dagger}{D^\mu}{\Sigma})
    +\frac{1}{4}{{f_\pi^2}{m_\pi^2}}{\rm Tr}
({\Sigma}+{\Sigma ^\dagger}) \nonumber \\
& & +i N^\dagger (v\cdot D) N
               + 2g_A N^\dagger (A\cdot S)N \nonumber \\
 & & -\frac{C_0}{8}\left[3 N^\dagger N \, N^\dagger N
                  +N^\dagger S N \cdot N^\dagger S N\right] +\ldots
\label{L0}
\end{eqnarray} 
with $g_A=1.267$ the axial-vector coupling of the nucleon and $C_0$ a
parameter determined from $NN$ scattering in the ${}^3$S$_1$ channel.
The next-to-leading-order (NLO) terms are contained in
\begin{eqnarray}
{\cal L}^{(1)}& = &
\frac{1}{2M}N^\dagger\left\{{-D^2}+(v\cdot D)^2
                    +2i{g_A}\{{v\cdot A},{S\cdot D}\} \right.\nonumber \\    
& &-\frac{i}{2}[{S^\mu},{S^\nu}]\left[(1+{\kappa _v}){f_{\mu\nu}}
+ \frac{1}{2}({\kappa_s}-{\kappa_v}){\rm Tr}{f_{\mu\nu}}\right] \nonumber \\
& &\left.+ 2M c_1 m_\pi^2 {\rm Tr}({\Sigma}+{\Sigma ^\dagger}) 
+ \left(8M c_2-g_A^2\right) (v\cdot A)^2
+ 8M c_3 A^2 \right\}N+\ldots,
\label{L1}
\end{eqnarray}
where ${\kappa _v}={\kappa _p}-{\kappa _n}$ and ${\kappa _s}={\kappa
_p}+{\kappa _n}$ are parameters related to the anomalous magnetic
moments of the proton, $\kappa _p=1.79$, and neutron, $\kappa
_n=-1.91$; and $c_1$, $c_2$, and $c_3$ are parameters that can
be determined from $\pi N$ scattering.
We use $c_1=-0.81$ GeV$^{-1}$, $c_2=2.5$ GeV$^{-1}$,
and $c_3=-3.8$ GeV$^{-1}$ \cite{fettes}
(compare with similar values obtained from $NN$ scattering \cite{Re03}).
The ellipses in Eq.~(\ref{L1}) indicate that we have not written down
pieces of ${\cal L}^{(1)}$ which are not relevant to our computation
of $\gamma N$ and $\gamma$d scattering. 

At sub-sub-leading order (N$^2$LO):
\begin{eqnarray}
{\cal L}^{(2)} &=& -\frac{e^2}{32 \pi^2 f_\pi} \pi_3
                  \epsilon^{\mu\nu\rho\sigma}F_{\mu\nu}F_{\rho\sigma}
\nonumber\\
&&-\frac{C_2}{8}
 \left[3N^\dagger N\, N^\dagger \left(-D^2 + (v\cdot D)^2\right)N
 +N^\dagger SN \cdot N^\dagger S \left(-D^2 +(v\cdot D)^2\right)N+h.c.\right]
\nonumber\\
&&-\frac{C_\epsilon}{2}
  \left[N^\dagger (S\cdot D) N \, N^\dagger (S \cdot D) N + h.c. \right]
+\ldots
\label{L2}
\end{eqnarray}
where the first term is from the chiral anomaly, with $\epsilon_{0123}=1$,
and the other two terms are corrections to the two-nucleon
potential in the deuteron channel, with the coefficients $C_2$ and
$C_\epsilon$ determined by fits to $NN$ scattering data.

In Eq.~(\ref{L2}) we do not explicitly display terms which are needed
for our computation below and have coefficients that are determined by
Lorentz invariance. These coefficients are dimensionless numbers times
$1/M^2$, and the relevant numbers can be extracted from
Ref.~\cite{Fe00}. A fixed-coefficient $\gamma N$ seagull from ${\cal
L}^{(2)}$ which enters the NLO $\gamma N$ amplitude stems from the
12th and 13th terms of Eq.~(3.8) in Ref.~\cite{Fe00}, while the $\gamma
NN$ pieces of ${\cal L}^{(2)}$ that contribute to $\gamma N$
scattering at N$^2$LO can be obtained from the 5th, 9th, 12th and 13th
terms of Eq~(3.8) and the last term of Eq.~(3.9) in that paper.

Finally, the relevant N$^3$LO terms are given by
\begin{equation}
{\cal L}^{(3)} = 2 \pi N^\dagger 
                 \left\{
  \left[\delta\beta_N + \delta\beta_v\tau_3\right] \smfrac{1}{2} g_{\mu\nu}
-    \left[(\delta\alpha_N + \delta\beta_N)
        + (\delta\alpha_v + \delta\beta_v)\tau_3\right]
   v_\mu v_\nu\right\}
F^{\mu\rho} F^{\nu}_{\; \; \rho} N +\ldots,
\label{L3}
\end{equation}
where $\delta\alpha_N$ ($\delta\beta_N$) and $\delta\alpha_v$
($\delta\beta_v$) are short-range contributions to the isoscalar and
isovector electric (magnetic) polarizabilities of the nucleon,
$\delta\alpha_p=(\delta\alpha_N+\delta\alpha_v)/2$ and
$\delta\alpha_n=(\delta\alpha_N-\delta\alpha_v)/2$ (similarly for
$\delta \beta_p$ and $\delta \beta_n$), which we seek to determine.
These LECs are linear combinations of $e_{89}$--$e_{94}$ of Fettes
{\it et al.}~\cite{Fe00}. The fixed-coefficient pieces of
$e_{89}$--$e_{94}$ are given in Table 6 of that paper.  Other relevant
terms in ${\cal L}^{(3)}$ which have only fixed coefficients and are
not written in Eq.~(\ref{L3}) but can be found in Ref.~\cite{Fe00} are
$O_{117}$ and $O_{118}$ of Table 5, as well as $X_{41}$ and $X_{53}$
of Table 8 and $Y_{11}$ of Table 9.

\subsection{Power counting}

\label{sec-pc}

Scattering amplitudes are computed from Feynman diagrams built from
the effective Lagrangian.  Observables are then power series in
momenta and pion masses, with the non-analyticities mandated by
perturbative unitarity.  The generic form for an amplitude is shown in
Eq. (\ref{pionfulT}).

Counting powers of $Q$ is simple, except for one important subtlety.
The simple part is that each derivative or power of the pion mass at a
vertex contributes $Q$ to the order of a diagram, each loop integral
contributes $Q^4$, each delta function of four-momentum
conservation contributes $Q^{-4}$, and each pion propagator contributes
$Q^{-2}$.  The complication arises from the fact that in multi-nucleon
systems there exist intermediate states that differ in energy from the
initial and final states by only the kinetic energy of nucleons, which
is $O(Q^2/M)$. Diagrams in which such intermediate states appear are
called {\it reducible}, while ones containing only intermediate
states that differ in energy from initial/final states by an amount
of $O(Q)$ are called {\it irreducible}.

First we consider only irreducible diagrams. By definition these
are graphs in which the energies flowing through all internal lines
are of $O(Q)$.  A nucleon propagator then contributes $Q^{-1}$, and we
find that a diagram with $A$ nucleons, $L$ loops, $C$ separately
connected pieces, and $V_i$ vertices of type $i$ contributes $Q^\nu$,
with~\cite{weinnp,friar}
\begin{equation}
\nu = 4-{A}-2C+2L+{\sum _i}{V_i}{\Delta _i}.
\label{nu}
\end{equation}
Eq. (\ref{pionfulT}) follows immediately
from the assumption that the parameters in
the effective Lagrangian scale as numbers of $O(1)$
times the appropriate power of $\lsc$.
Hence the leading {\it
irreducible} graphs are tree graphs ($L=0$) with the maximum number of
separately connected pieces ($C=max$), and with 
leading-order ($\Delta_i =0$) vertices.  
Higher-order graphs (with larger powers of $\nu$) are
perturbative corrections to this leading effect. In this way, a
perturbative series is generated by increasing the number of loops
$L$, decreasing the number of connected pieces $C$, and increasing the
number of derivatives, pion masses and/or nucleon fields in
interactions.

Processes involving one or zero nucleons and any number of light
particles with $p\sim m_\pi$ only receive contributions from
irreducible diagrams, and there is good evidence that the resulting
\cpt~power counting defined by Eq.~(\ref{nu}) works in most kinematics
(exceptions are possible, e.g., near threshold pion three-momenta
may be smaller than the $\sim m_\pi$ assumed in deriving
Eq.~(\ref{nu}))~\cite{bkmrev}.  In processes involving more than one
nucleon, the full amplitude consists of irreducible subdiagrams linked
by reducible states.  New divergences arise from these reducible
intermediate states, which can potentially upset the scaling of the
EFT parameters with $\lsc$.  Weinberg \cite{weinnp} has implicitly
assumed that this is not the case, and thus that the \cpt power
counting continues to hold for irreducible diagrams.

It is now known that this assumption is not always correct
\cite{Ka96}.  In particular, consistency of renormalization in
$NN$ scattering seems to require that some pion-mass
contributions should be demoted to higher orders than in Weinberg's
original power counting. On the other hand, the error of not doing so
is apparently numerically small~\cite{chiralexp}. This explains the
phenomenological successes of applications of Weinberg's
proposal to few-nucleon systems \cite{bkreview,friar},
including to $\gamma$d scattering~\cite{therestofus} and a 
number of other reactions involving external probes of
deuterium~\cite{hybrid}. Here we continue to organize our calculation
according to the power counting (\ref{nu}), but we will see some indications
that the irreducible diagrams for Compton scattering on deuterium are
not behaving exactly as Weinberg's straightforward chiral power
counting for them would suggest.

In an $A$-nucleon system which is not subject to any external probes,
the sum of $A$-nucleon irreducible diagrams is, by definition, the
potential $V$.  To obtain the full two-nucleon Green's function $G$,
these irreducible graphs are iterated using the free two-nucleon
Green's function $G_0$, which can, of course, have a small energy
denominator that does not obey the \cpt power-counting: $G=G_0 + G_0 V
G$. Making a spectral decomposition of $G$ we can extract the piece
corresponding to the deuteron pole.  Since $V$ is two-particle
irreducible we may perform a chiral power counting on it as per
Eq. (\ref{nu}).  Such a potential was constructed including terms up
to $\nu =3$ and with an explicit Delta field in
Refs.~\cite{ordonez}.  Better fits have now been achieved using a
potential without explicit Delta's at $\nu =3$~\cite{Ep99,EM02}.
Recently the first calculation of $V$ to $\nu=4$ appeared, and a good
fit to $NN$ scattering data for lab energies below 290 MeV
resulted~\cite{EM03}.  Here we use the wave function obtained from the
NLO ($\nu=2$) potential of Ref.~\cite{Ep99}, as that is the consistent
choice given the chiral order at which we calculate $\gamma$d
scattering.  Since this is only an NLO potential the quality of the
fit to $NN$ scattering data is not nearly as good as that found in
modern phenomenological potentials, or in Ref.~\cite{EM03}. In order
to get a sense of how large the impact of higher-order terms in $V$
might be, we have also computed $\gamma$d scattering using deuteron
wave functions generated from a modern one-boson-exchange $NN$
potential \cite{Nijm93}, and the Bonn OBEPQ~\cite{Bonnrep}.

When we consider external probes with momenta of order $Q$, the sum of
irreducible diagrams forms a kernel $K$ for the process of interest,
here the kernel $K_{\gamma \gamma}$ for $\gamma NN \rightarrow \gamma
NN$.  Again, the power counting of Eq.~(\ref{nu}) applies to this
object, since all two-nucleon propagators in it scale as $Q^{-1}$, as
long as the photon energy $\omega \sim Q$.  Note also that in order to
have an irreducible graph in which two photons are attached to
different nucleons the two nucleons must be connected by some
interaction, so that there is a mechanism to carry the energy of order
$Q$ from one nucleon to the other. Thus such graphs should be regarded
as having one separately connected piece ($C=1$), so that the dictates
of four-momentum conservation can be respected.

The full Green's function $G_{\gamma \gamma}$ for the reaction is now
found by multiplying the kernel $K_{\gamma \gamma}$ by two-particle
Green's functions in which the $NN$ propagators with small
energy denominators may appear, $G_{\gamma \gamma}=G K_{\gamma \gamma}
G$.  Extracting the piece of $G_{\gamma \gamma}$ corresponding to the
deuteron pole at $E=-B$ in the both the initial and final state we
obtain the $\gamma$d $T$-matrix:
\begin{equation}
T = \langle\psi|K_{\gamma\gamma}|\psi\rangle,
\label{nuclearT}
\end{equation}
with $|\psi \rangle$, the deuteron wave function, being the solution to:
\begin{equation}
\left(\frac{\hat{p}^2}{M} + V\right)|\psi \rangle=-B|\psi \rangle.
\end{equation}
Full consistency can be achieved by treating both the kernel for the
interaction with external probe and the $NN$ potential $V$ in \cpt and
expanding both to the same order in $\nu$.  Note that if $K_{\gamma
\gamma}$ is constructed in a gauge invariant way and is worked out to
the same order in \cpt as $V$ then this calculation will be gauge
invariant up to that order in $\chi$PT.  It will not, however, be {\it
exactly} gauge invariant.  Contributions to $K_{\gamma \gamma}$ of the
form $K_\gamma^\dagger G K_\gamma$ (with $K_\gamma$ the
two-nucleon-irreducible kernel for one photon interacting with the
$NN$ system) only appear as part of $K_{\gamma \gamma}$
order-by-order in the expansion in $Q/\lsc$. For gauge invariance to
be exact this contribution from two-nucleon intermediate states must be
included in full in $G_{\gamma \gamma}$.

This is not a deep problem.  In the very-low-energy region $\omega
\sim m_\pi^2/M$, the power-counting of Eq.~(\ref{nu}) does not apply
to $K_{\gamma \gamma}$, since graphs in $K_{\gamma \gamma}$ with
two-nucleon intermediate states will have denominators of order
$Q^2/M$.  There are thus {\it two} regimes for Compton scattering on
nuclei, depending on the relation of the energy of the photonic probe,
$\omega$, to the typical nuclear binding scale $\sim
m_\pi^2/M$. (Compare with the discussion of low-energy theorems for
threshold pion photoproduction on nuclei in Ref. \cite{FT81}, and of
Compton scattering on the deuteron with perturbative pions in
Ref.~\cite{Ch98}.)  Here we are interested mostly in photon energies
between 60 and 200 MeV, and so we regard ourselves as being firmly in
the regime $\omega \sim m_\pi \gg m_\pi^2/M$.  In this regime it is
valid to treat the contributions $K_\gamma^\dagger G K_\gamma$
perturbatively in the $Q$ expansion, i.e. we can use a chiral
expansion for the two-nucleon, fully-interacting, $G$ which appears in
$K_\gamma^\dagger G K_\gamma$. This results in a calculation that is
completely gauge invariant up to the order in the chiral expansion to
which we work. However, such a treatment fails once $\omega \sim
m_\pi^2/M$, and so---as will become manifest below---our conclusions regarding
$\gamma$d scattering at these lower energies should be viewed with
some caution. In this very-low-energy domain it seems sensible to
integrate out the pion, and then analyze mechanisms and count powers
in the ``pionless'' EFT~\cite{griess}.

\subsection{Notation}

\label{sec-not}

A word of explanation about notation is required.  We are interested
here in Compton scattering on both the nucleon and the deuteron to 
next-to-next-to-leading order, since it is at that order that
short-distance contributions to the polarizabilities first appear.
These short-range $\gamma N$ interactions have index $\Delta=3$, see
Eq. (\ref{L3}).  On the other hand, the corresponding value of the
index $\nu$ defined in Eq.~(\ref{nu}) depends on the target.

For scattering on a single nucleon, $A=1$ and $C=1$ in Eq. (\ref{nu}).
The first non-vanishing contribution to the $T$-matrix---the Thomson
term---appears at $\nu=2$, while
the short-distance polarizability terms
first show up at $\nu=4$.
One-nucleon
scattering diagrams are traditionally labeled as $Q^\nu$. 
The highest order we work to here is thus $O(Q^4)$. 

For scattering on the deuteron, $A=2$ and $C=1,2$ in Eq. (\ref{nu}).
The contributions of the single-nucleon amplitude to $K_{\gamma\gamma}$
have $C=2$. The single-nucleon Thomson term
therefore has $\nu=-1$, while the short-distance polarizability terms first
appear at $\nu=1$.
However, in order to make closer contact with previous work on 
$\gamma N$ scattering in \cpt~\cite{ulf1,ulf2,judith}, 
we follow the single-nucleon practice
and refer to our $\nu=1$ deuteron calculation as being  $O(Q^4)$. 
Of course only relative orders are physically significant.

A final warning. Our index $\Delta$ counts simultaneously the number
of derivatives and nucleon fields.  Sometimes the EFT Lagrangian is
broken down into a purely-pionic Lagrangian, a pion-nucleon
Lagrangian, and so on, and the terms in each are further labeled not
by $\Delta$ but by the number of derivatives.  For example, our ${\cal
L}^{(0)}$ of Eq.~(\ref{L0}) is often written as ${\cal
L}_{\pi\pi}^{(2)}+{\cal L}_{\pi N}^{(1)}+{\cal L}_{NN}^{(0)}$.  These
notational issues should be borne in mind when referring to
Refs.~\cite{judith,therestofus}.

\section{Compton scattering on the proton: theory}
\label{sec-prottheory}

In this section we consider Compton scattering on a single nucleon.
This is required for the extraction of the proton polarizabilities
from proton Compton scattering data, and also because the \cpt $\gamma
N$ amplitude is an important ingredient in our calculation of
$\gamma$d scattering.  We work in the Coulomb gauge where the incoming
and outgoing photon momenta $k=(\omega, \vec{k})$ and $k'=(\omega',
\vec{k}')$ and polarization vectors $\epsilon$ and $\epsilon'$
satisfy:
\begin{equation}
k^2=0; \qquad k'^2=0; \qquad v \cdot \epsilon=0; \qquad v \cdot \epsilon' =0.
\end{equation}
 
In either the Breit frame or the $\gamma N$ center-of-mass frame
the Compton scattering amplitude can be written as:
\begin{eqnarray}
&&T_{\gamma N}= \left\{ A_1 \vepsprime\cdot\veps
            +A_2 \vepsprime\cdot \hat{k} \, \veps\cdot \hat{k'}
             +iA_3 \vsigma\cdot (\vepsprime\times\veps)
             +iA_4 \vsigma\cdot (\hat{k'}\times\hat{k})\, \vepsprime\cdot\veps
     \right. \nonumber \\
 & & \left.
\!\!\!\!\!\!\!\!\!
     +iA_5 \vsigma\cdot [(\vepsprime\times\hat{k})\, \veps\cdot\hat{k'}
            -(\veps\times\hat{k'})\, \vepsprime\cdot\hat{k}]
     +iA_6 \vsigma\cdot [(\vepsprime\times\hat{k'})\, \veps\cdot\hat{k'}
                        -(\veps\times\hat{k})\, \vepsprime\cdot\hat{k}] \right\},
\label{eq:Ti}
\end{eqnarray}
with $A_i$, $i=1\ldots 6$ scalar functions of photon energy and scattering
angle.

\begin{figure}[tb]
  \begin{center} \mbox{\epsfig{file=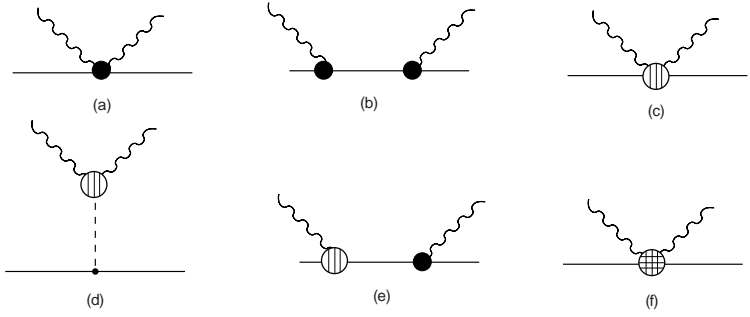,width=10truecm,angle=0}}
  \end{center}
\caption{\label{protborn} Tree diagrams that contribute to Compton
scattering in the $\epsilon\cdot v=0$ gauge up to $O(Q^4)$. Small dots
are vertices from ${\cal L}^{(0)}$, larger dots are vertices from
${\cal L}^{(1)}$, the sliced dots are vertices from ${\cal L}^{(2)}$
and the sliced and diced dots are vertices from ${\cal
L}^{(3)}$. Crossed graphs and graphs which differ only in the ordering
of vertices from ${\cal L}^{(1)}$ and ${\cal L}^{(2)}$ are included in
the calculation, but are not shown here.}
\end{figure}

In the relativistic theory the Compton scattering amplitude can be
split into Born and non-Born---or ``structure''---terms.  The six
polarizabilities $\alpha$, $\beta$, and $\gamma_1$--$\gamma_4$ are the
independent coefficients that arise when these structure terms are expanded
in powers of $\omega$ up to $\omega^3$.  For the Breit-frame amplitude
such an $\omega$-expansion gives (keeping terms up to $O(1/\mn^3)$):
\begin{eqnarray}
A_1(\omega,\theta)\!&=&\!-{\Z^2e^2\over\mn}+{e^2\over 4\mn^3}
\Bigl((\Z+\kappa)^2(1+\cos\theta)-\Z^2\Bigr)(1-\cos\theta)\,\omega^2\nonumber\\
&&\qquad \qquad \qquad \qquad + 4\pi(\alpha + \beta \, \cos\theta)\omega^2+
O(\omega^4)\nonumber\\ 
A_2(\omega,\theta)\!&=&\!{e^2\over
4\mn^3}\kappa(2\Z+\kappa) \omega^2 \cos\theta -4\pi\beta \omega^2 \,+
O(\omega^4)\nonumber\\ 
A_3(\omega,\theta)\!&=&\!  {e^2 \omega\over
2\mn^2}\Bigl(\Z(\Z+2\kappa)-(\Z+\kappa)^2 \cos\theta\Bigr)
+A_3^{\pi^0}\nonumber\\
&& \qquad \qquad \qquad \qquad 
+ 4\pi\omega^3(\gamma_1 - (\gamma_2 + 2 \gamma_4) \, \cos\theta)+
O(\omega^5)\nonumber\\ A_4(\omega,\theta)\!&=&\! -{e^2 \omega \over 2
\mn^2 }(\Z+\kappa)^2 + A_4^{\pi^0}+4\pi\omega^3 \gamma_2 +
O(\omega^5)\nonumber\\ A_5(\omega,\theta)\!&=&\! {e^2 \omega \over
2 \mn^2 }(\Z+\kappa)^2 + A_5^{\pi^0}+4\pi\omega^3\gamma_4 +
O(\omega^5)\nonumber\\ A_6(\omega,\theta)\!&=&\! -{e^2 \omega \over
2 \mn^2 }\Z(\Z+\kappa) +A_6^{\pi^0}+4\pi\omega^3\gamma_3 +
O(\omega^5),
\label{ampexp}
\end{eqnarray}
with $A_3^{\pi^0}$--$A_6^{\pi^0}$ the contributions from the pion-pole
graph, Fig.~\ref{protborn}(d).

\begin{figure}[tb]
  \begin{center} \mbox{\epsfig{file=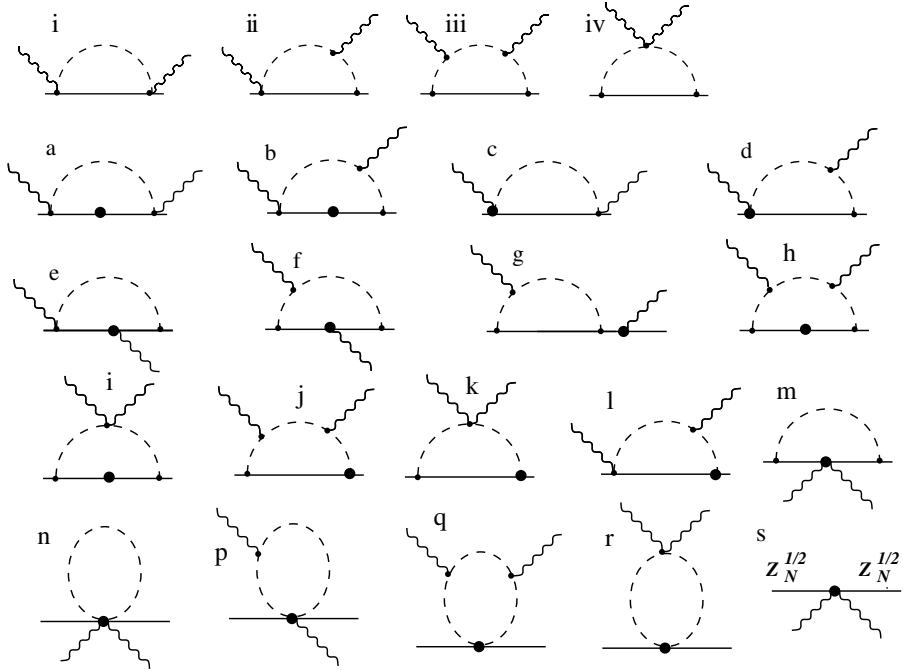,width=12truecm,angle=0}}
  \end{center}
\vskip -1.2cm
\caption{\label{protloop} Diagrams which contribute to nucleon Compton
scattering in the $\epsilon\cdot v=0$ gauge at 3rd (i-iv) and 4th
(a-s) order. Vertices are labeled as in Fig.~\ref{protborn}.}
\end{figure}

In heavy-baryon chiral perturbation theory the amplitudes $A_1$--$A_6$
include the tree contributions shown in Fig.~\ref{protborn}, and the
pion-loop contributions of Fig.~\ref{protloop}. At $O(Q^2)$ only graph
\ref{protborn}(a) contributes, and this reproduces the
spin-independent low-energy theorem for Compton scattering from a
single nucleon---the Thomson limit:
\begin{equation}
T_{\gamma N}=-\veps \cdot \vepsprime \frac{{\cal Z}^2 e^2}{M},
\label{eq:OQ2}
\end{equation}
where ${{\cal Z}}$ is the nucleon charge in units of $|e|$.  

At $O(Q^3)$ the s-channel proton-pole diagram, Fig.~\ref{protborn}(b),
and its crossed u-channel partner, together with a $\gamma N$ seagull
from ${\cal L}^{(2)}$ (Fig.~\ref{protborn}(c)) ensure that HB$\chi$PT
recovers the Low, Gell-Mann and Goldberger low-energy theorem for
spin-dependent Compton scattering \cite{LGG}.  At $O(Q^3)$ the
t-channel pion-pole graph Fig.~\ref{protborn}(d) also enters; its
contribution, which varies rapidly with energy, is often included in
the definition of the spin polarizabilities and is the reason that the
backward spin polarizability is so much larger in magnitude than the
forward one.  Pion-loop graphs enter first at $O(Q^3)$---
see graphs i-iv of Fig.~\ref{protloop}---and give
energy-dependent contributions to the amplitude which include the
well-known predictions for the spin-independent polarizabilities 
of Eq.~(\ref{eq:betaOQ3}), as well as less-famous predictions for 
$\gamma_1$--$\gamma_4$~\cite{ulf1}. The third-order loop amplitude of
Refs.~\cite{bkmrev,judith} is obtained from diagrams
\ref{protloop}(i)--\ref{protloop}(iv)---together with graphs related
to them by crossing and alternative time-ordering of vertices---and is
given in Appendix~\ref{ap-oq3prot}. As observed in
Ref.~\cite{babusci}, the $O(Q^3)$ \cpt result for $\gamma$p scattering
is in good agreement with the data at forward angles up to quite high
energies ($\sim 200$ MeV), but fails to describe the backward-angle
data already at around 100 MeV.

At $O(Q^4)$ in HB$\chi$PT nucleon pole graphs \ref{protborn}(e) and
the fixed-coefficient piece of the seagull depicted in
Fig.~\ref{protborn}(f) give further terms in the $1/M$-expansion of
the relativistic Born amplitude. There are also a number of
fourth-order one-pion loop graphs (graphs a-r of Fig.~\ref{protloop}),
which---with one exception---contribute only to the ``structure'' part
of the $\gamma N$ amplitude.  (The exception is graph \ref{protloop}g
which also renormalizes $\kappa$ in the spin-dependent Born amplitude.) All
of the graphs \ref{protloop}a--\ref{protloop}r involve vertices from
${\cal L}^{(1)}$ of Eq.~(\ref{L1}). In particular, the expressions for
diagrams \ref{protloop}n--\ref{protloop}r contain the LECs $c_1$,
$c_2$, and $c_3$,
and in diagrams \ref{protloop}e--\ref{protloop}g 
the nucleon
anomalous magnetic moments enter. 

Divergences in graphs \ref{protloop}a--\ref{protloop}f, 
\ref{protloop}h--\ref{protloop}k, \ref{protloop}p, and \ref{protloop}q 
need to be
renormalized by the counterterms appearing in Eq.~(\ref{L3}).  The
divergent fourth-order loop contributions to the spin-independent
polarizabilities---first derived in Ref.~\cite{ulf2}---are as 
follows:
\begin{eqnarray}
&&(\alpha+\beta)^{(4)}_{\rm loop} \, \, : \, \, {e^2\over 4 \pi \fpi^2}\left({g_A^2\over 12 \mn}\Bigl(
(d^2+8d-1)+(12\mu_s+2d+4)\tau_3\Bigr)-\frac 2 3 \tilde c_2\right)
\Delta_\pi';\nonumber\\&&\beta^{(4)}_{\rm loop} \, \, : \, \, {e^2\over
4 \pi \fpi^2}\left({g_A^2\over 24 \mn} \Bigl(-d^2+10d+23+24\mu_s\tau_3\Bigr)
-\frac 1 3 \Bigl(\tilde
c_2+(d-4)(c_3-2c_1)\Bigr)\right)\Delta_\pi',\nonumber\\
&&
\end{eqnarray}
where $d$ is the space-time dimension,
\begin{equation}
\Delta_\pi'=(d-2)L(\mu)+(1/8\pi^2)\log\left({m_\pi\over\mu}\right),
\end{equation}
$\mu$ is the renormalization scale, and
\begin{equation}
(d-4)L(\mu)=(1/16\pi^2) +O(d-4).
\end{equation}

Now adding the counterterms from ${\cal L}^{(3)}$ of Eq.~(\ref{L3})
cancels the divergences, leaving:
\begin{eqnarray}
4\pi(\alpha+\beta)^{(4)}&=&{e^2\over
16\pi^2\fpi^2}\left({g_A^2\over 12 \mn}\Bigl(
(94+24(\mu_s+1)\tau_3)\log\left({m_\pi \over \mu}\right)+79+(16+12\mu_s)\tau_3\Bigr)
\right.\nonumber\\&&\left.  -\frac 2 3 \tilde
c_2\Bigl(2\log\left({m_\pi \over \mu}\right)+1\Bigr)\right)
+ \delta \alpha_N(\mu) + \delta \beta_N(\mu)
+ \tau_3 \left(\delta \alpha_v(\mu) + 
\delta \beta_v(\mu)\right); \nonumber\\
4\pi\beta^{(4)}&=&{e^2\over
16\pi^2\fpi^2}\left({g_A^2\over 24 \mn}\Bigl(
(94+48\mu_s\tau_3)\log\left({m_\pi \over \mu}\right)+51+24\mu_s\tau_3 \Bigr)
\right.\nonumber\\&&\qquad \quad \left.  -\frac 2 3 \tilde
c_2 \log\left({m_\pi \over \mu}\right)-\frac 1 3 (\tilde c_2+2c_3-4c_1)\right)
+ \delta \beta_N(\mu) + \tau_3 \delta \beta_v(\mu),\nonumber\\
\label{eq:fourthorderpols}
\end{eqnarray}
with $\tilde c_2=c_2 - \frac{g_A^2}{8 M}$.

The scale dependence in the LECs $\delta \alpha_{N,v}$ and $\delta
\beta_{N,v}$ cancels the $\mu$-dependence of the loop pieces. The
$\gamma N$ seagulls proportional to these LECs
(Fig.~\ref{protborn}(f)) represent contributions to $\alpha$ and
$\beta$ from mechanisms other than soft-pion loops,
i.e. short-distance effects.  As such they should, strictly speaking,
be determined from data, since they contain physics that is outside
the purview of the EFT.  In contrast the spin polarizabilities, as
well as all higher terms in the Taylor-series expansion of the Compton
amplitude, are still predictive in $\chi$PT at this
order---as long as the values of the $c_i$'s are taken from some other
process. The first short-distance effects in $\gamma_1$--$\gamma_4$
occur at $O(Q^5)$.

The full amplitudes at $O(Q^4)$ are too lengthy to be given here; they
can be constructed from the information given in the appendix of
Ref.~\cite{judith} and the details given in Appendices
\ref{ap-oq4prot} and \ref{ap-borngen}. We stress that in what follows
it is always the full $O(Q^4)$ HB$\chi$PT amplitude, and not a
Taylor-series expansion such as (\ref{ampexp}), which is employed.

A well-known problem with the heavy-baryon approach is that at finite
order the pion-production threshold comes at $\omega=m_\pi$.
Furthermore beyond third order the amplitudes diverge at that point.
The solution is to resum the terms which, taken together to all
orders, would shift the threshold to the physical point $\omega_{\rm
th}$; in practice this involves writing the amplitudes as functions of
$\omega/\omega_{\rm th}$ instead of $\omega/\mpi$, then
Taylor-expanding to correct the error introduced to appropriate order.
The resulting amplitudes are then finite and have the threshold in the
correct place.  This procedure was first introduced for forward
scattering in Ref.~\cite{ulf1}, and elaborated for non-forward
scattering in Ref.~\cite{judith}.  Since the proton data fitted spans
the pion-production threshold we used this procedure for the
extraction of the proton polarizabilities, but as the highest energy
in the deuteron case is only 95~MeV, we did not do so in that case.

Having obtained the amplitudes, the differential
cross section in any frame is a kinematic prefactor $\Phi^2$ times the
invariant  $|T|^2$. 
Experimental data, though taken
in the lab frame, are sometimes presented as c.m.-frame
differential cross sections, so both prefactors are needed:
\begin{equation} \Phi_{\rm lab}={1\over 4\pi}{E_\gamma'\over
E_\gamma};\qquad\qquad  \Phi_{\rm  cm}={1\over  4\pi}{\mn\over\sqrt{s}}.
\end{equation}
Finally, in terms of the Breit-frame $A_1$--$A_6$, $|T|^2$ is given by:
\begin{eqnarray}
|T|^2&=& \half A_1^2  ( 1 + \cos^2\theta) + \half A_3^2
(3-\cos^2\theta) \nonumber \\ 
&& + \sin^2 \theta \, \bigl[ 4 A_3 A_6  +( A_3 A_4 +
2 A_3 A_5 -A_1 A_2)\cos \theta\bigr]\nonumber \\ 
&& + \sin^2 \theta \,\bigl[\half
A_2^2 \sin^2 \theta + \half A_4^2(1+\cos^2 \theta) + A_5^2
(1+2\cos^2 \theta) +3A_6^2 \nonumber \\ 
&& \kern2cm+ 2A_6 (A_4 + 3 A_5)\cos \theta + 2 A_4 A_5
\cos^2 \theta\bigr].
\label{tsqrd}
\end{eqnarray}
(Above the $\pi N$ threshold one should replace $A_i^2$ by $ |A_i|^2$
and $A_i A_j$ by $ \Re(A_i^* A_j)$.)  

\section{Compton scattering on the proton: results}

\label{sec-protres}

The only unknowns in the $O(Q^4)$ \cpt $\gamma$p amplitude are $\delta
\alpha_p$ and $\delta \beta_p$, since the values of $c_1$, $c_2$, and
$c_3$ can be determined from $\pi N$~\cite{fettes} (or
$NN$~\cite{Re03}) scattering data. If these counterterms are combined
with the (scale-dependent) divergences they renormalize we can regard
$\alpha_p$ and $\beta_p$ (or, more precisely, the shifts of the
polarizabilities from their $O(Q^3)$ values) as the two free
parameters in the $O(Q^4)$ calculation.

In Ref.~\cite{judith}, these two free parameters were not fit
directly.  Rather, the input to that calculation included the Particle
Data Group (PDG) values of the polarizabilities \cite{PDG}. The resulting
differential cross sections are in good agreement with the low-energy
data.

While these PDG values are derived from polarizabilities quoted in
experimental papers, the energies of the relevant experiments are, for
the most part, too high for a direct fit to the polynomial form
(\ref{ampexp}). Instead the amplitudes these papers employ in order
to extract $\alpha_p$ and $\beta_p$ are based on dispersion relations,
using plausible assumptions about their asymptotic form coupled with
models for those amplitudes which do not converge.  The difference
$\alpha_p - \beta_p$ is a parameter in these fits; the sum is
(usually) taken from the Baldin sum rule~\cite{baldin}.

This approach implicitly builds in model-dependent assumptions about
the behavior of the $\gamma$p amplitude beyond the low-energy
regime. In what follows we adopt a different strategy, using \cpt to
determine the polarizabilities from only low-energy ($\omega \leq 200$
MeV) differential cross sections, so as to obtain truly
model-independent results for $\alpha_p$ and $\beta_p$.

In a recent paper the world data on proton Compton scattering was
examined, and it was demonstrated that data from the 1950s and 1960s
is compatible with the more modern data from 1974 on, and is useful in
reducing errors~\cite{Baranov}.  We have therefore used all the data
in Table 2 of Ref.~\cite{Baranov}, with the exception of the first
three points for which the normalization error is not known.  We have
also included higher-energy points from the experiments of
Refs.~\cite{pdata}, and the TAPS data~\cite{mainz}.

All the experiments have an overall normalization error.  We incorporate this 
by adding a piece to the usual $\chi^2$ function:
\begin{eqnarray}
\chi^2&=&\chi^2_{\rm stat.}+\chi^2_{\rm syst.}\nonumber\\
&=&\sum_{j=1}^{N_{\rm sets}}\sum_{i=1}^{N_j}
\left(\frac{ {\cal N}_j (\d\sigma_{ij}/\d\Omega)_{\rm expt} -(\d\sigma_{ij}/\d\Omega)_{\rm theory}} {{\cal N}_j \Delta_{ij}}\right)^2
+\sum_{j=1}^{N_{\rm sets}}
\left(\frac{{\cal N}_j - 1}{{\cal N}_j \delta_j}\right)^2,
\label{eq:chisqare}
\end{eqnarray}
where $\d\sigma_{ij}$ and $\Delta_{ij}$ are the value and statistical
error of the $i$th observation from the $j$th experimental set,
$\delta_j$ is the fractional systematic error of set $j$, and ${\cal
N}_j$ is an overall normalization for set $j$.  The additional
parameters ${\cal N}_j$ are to be optimized by minimizing the combined
$\chi^2$.  This can be done analytically, leaving a $\chi^2$ that is a
function of $\alpha_p$ and $\beta_p$ alone.  Since we fit 13 $\gamma$p
data sets this reduces the number of degrees of freedom in the final
minimization by 13.
(For more details on this formalism see Ref.~\cite{Baranov}.)

In Ref.~\cite{judith} it was found that while the \cpt fit was very
good at sufficiently low energies and forward angles, it failed for
higher energies and backward angles, and did so in a way that could
not be remedied by adjusting parameters. This suggests that at a
relatively low momentum ``new'' physics is coming in---an observation
that is consistent with the Delta-nucleon mass difference being only a
factor of two larger than $m_\pi$.  Indeed, in both existing
models and the effective field theory with explicit Delta-isobar
degrees of freedom~\cite{PP03,Hi04} Delta effects appear at backward
angles already at $\omega \sim m_\pi$. (Their impact on the
forward-angle data does not show up until higher energies are
reached.)  As a consequence in our extraction of $\alpha_p$ and
$\beta_p$ we impose a cut on both the photon energy and the momentum
transfer, $\omega,\, \sqrt{|t|}\le \omega_{\rm max}$.

\begin{figure}[htb]
   \epsfxsize=12.0cm
   \centerline{\epsffile{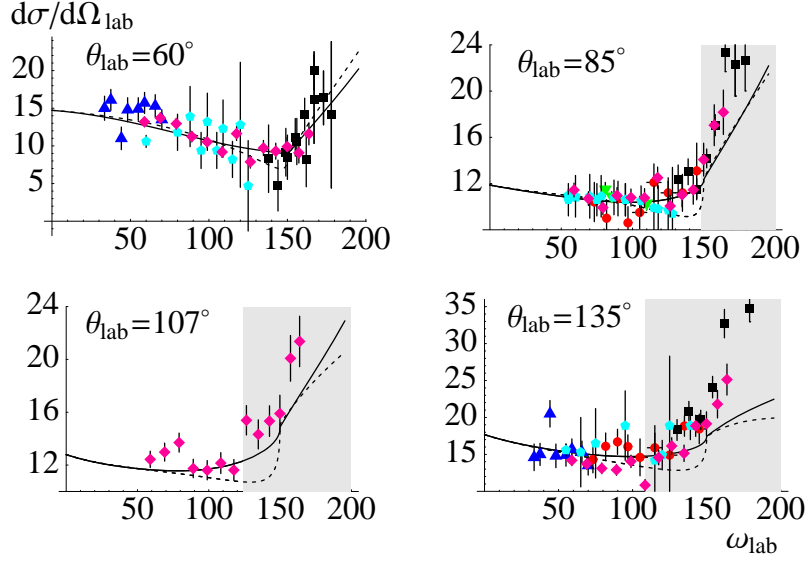}}
\caption{\label{Q4deuprotfit} Results of the $O(Q^4)$ EFT best fit to
   the differential cross sections for Compton scattering on the
   proton at various angles, compared to the experimental data
   \protect\cite{pdata,mainz}. The gray region is excluded from the
   fit. The magenta diamonds are Mainz data \cite{mainz}; the other
   symbols are explained in Ref. \cite{judith}.}
\end{figure}

We investigated the effect of varying $\omega_{\rm max}$ between 140
and 200~MeV---see Table~\ref{table-gampfits}. As $\omega_{\rm max}$
was increased the central values of $\alpha_p$ and $\beta_p$ changed
very little; the 1-$\sigma$ curve tightened as more data were
included, but the $\chi^2 /{\rm d.o.f}$ of the best-fit point tended
to rise.  This did not seem to be due to a systematic worsening of the
fit, but rather to the inclusion of rather noisy data.  We have chosen
to quote the central value corresponding to a cut-off of $\omega_{\rm
max}=180$~MeV, for which $\chi^2 /{\rm d.o.f}$ was (locally) minimized
at $133/113$.  This fit is shown in Fig. \ref{Q4deuprotfit} together
with the low-energy data.

\begin{table*}[tbp]
\begin{center}
\begin{tabular}{|c|c||c|c|c|}
\hline
$\omega_{\rm max}$ (MeV) & $N_{\rm data}$ & $\alpha_p$ ($10^{-4}$ fm$^3$) &
$\beta_p$ ($10^{-4}$ fm$^3$)& 
$\chi^2$/d.o.f.\\
\hline \hline
200 & 144 & 11.7 & 3.40 & 1.32\\
180 & 126 & 12.1 & 3.38 & 1.18\\
165 & 109 & 12.3 & 3.37 & 1.27\\
140 & 73  & 12.0 & 3.13 & 1.26\\
\hline
\end{tabular}
\caption{Results of $O(Q^4)$ fits for $\alpha_p$ and $\beta_p$ to
world $\gamma$p data over various ranges $\omega,\sqrt{|t|} <
\omega_{\rm max}$. In each case $N_{\rm data}$ is the number of data
points included in the fit. }
\label{table-gampfits}
\end{center}
\end{table*}

\begin{figure}[htb]
   \epsfxsize=10.0cm
  \centerline{\epsffile{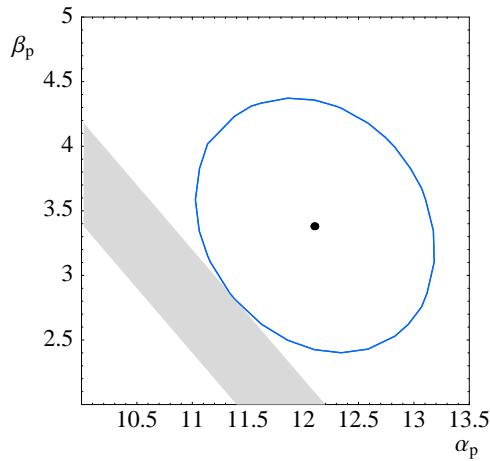}}
\caption{\label{alphabetap} Values (in $10^{-4} \, {\rm fm}^3$) of the
   proton magnetic and electric polarizabilities extracted from an
   $O(Q^4)$ EFT fit to low-energy cross-section data, together with
   the values allowed by the Baldin sum-rule constraint
   (\ref{eq:Baldin}) (gray shaded area).}
\end{figure}

The best-fit point for the proton electric and
magnetic polarizabilities is shown in Fig. \ref{alphabetap}, together
with the 1-$\sigma$ curve ($\chi^2_{\rm min}+2.28$):
\begin{eqnarray}
\alpha_p &=& (12.1 \pm 1.1)_{-0.5}^{+0.5} \times 10^{-4} \, {\rm fm}^3, 
\nonumber\\
\beta_p &=& (3.4 \pm 1.1)_{-0.1}^{+0.1} \times 10^{-4} \, {\rm fm}^3. 
\label{eq:protpol1}
\end{eqnarray}
Statistical (1-$\sigma$) errors are inside the brackets. An estimate
of the theoretical error due to truncation of the expansion at
$O(Q^4)$ is given outside the brackets. The result
(\ref{eq:protpol1}) is fully compatible with other extractions,
although the central value of $\beta_p$ is somewhat
phigher~\cite{PDG,drechsel}.

This theoretical error is assessed by varying the bound on which data
are fit from 140 MeV to 200 MeV.  The size inferred thereby is
consistent with estimates of the impact that $O(Q^5)$ terms would have
on the fit.  In this context it is worth noting that although the
$O(Q^4)$ $\chi$PT predictions for the spin polarizabilities
$\gamma_1$--$\gamma_4$ are not in good agreement with those obtained
from other sources, in Ref.~\cite{judith} it was shown that at the
energies considered here the differential cross section is largely
insensitive to their values.  (Asymmetries in polarised Compton
scattering would be far more sensitive.)

The recent re-evaluation of the Baldin sum rule~\cite{baldin} by Olmos
de Le\'on {\it et al.}~\cite{mainz} gives:
\begin{equation}
\alpha_p + \beta_p=(13.8 \pm 0.4) \times 10^{-4}~{\rm fm}^3.
\label{eq:Baldin}
\end{equation}
If one considers only statistical errors the results
(\ref{eq:protpol1}) and (\ref{eq:Baldin}) are marginally inconsistent
at the 1-$\sigma$ level---see Fig.~\ref{alphabetap}. Our values for
$\alpha_p$ and $\beta_p$ are consistent with the constraint
(\ref{eq:Baldin}) once the potential impact of theoretical
uncertainties is taken into account.

Including the constraint (\ref{eq:Baldin}) in our
fit leads to values for $\alpha_p$ and $\beta_p$ consistent with
Eq. (\ref{eq:protpol1}), but with smaller statistical errors, namely:
\begin{eqnarray}
\alpha_p &=& (11.0 \pm 0.5 \pm 0.2)_{-0.5}^{+0.5} \times 10^{-4} \, {\rm fm}^3,
\nonumber\\
\beta_p &=& (2.8 \pm 0.5 \mp 0.2)_{-0.1}^{+0.1} \times 10^{-4} \, {\rm fm}^3.
\label{eq:protpol2}
\end{eqnarray}
In Eq. (\ref{eq:protpol2}) we have left the systematic error unchanged,
but have now included a second error inside the brackets, whose source
is the error on the sum-rule evaluation (\ref{eq:Baldin}). The smaller
statistical errors are achieved at the expense of introducing
certain assumptions about the high-energy behavior of the Compton
amplitude into the result, via the use of the Baldin sum-rule
result (\ref{eq:Baldin}). The result (\ref{eq:protpol2}) is in
excellent agreement with a recent determination of $\alpha_p$ 
and $\beta_p$ which used an effective field
theory with explicit Delta degrees of freedom to fit all
$\gamma$p scattering data up to $\omega \sim 170$ MeV~\cite{Hi03}. 

\section{Compton scattering on the deuteron: theory}
\label{sec-deuttheory}

In this section we concentrate on real Compton scattering on an
$A=2$ system. 

\begin{figure}[thbp]
   \epsfysize=8.0cm
   \centerline{\epsffile{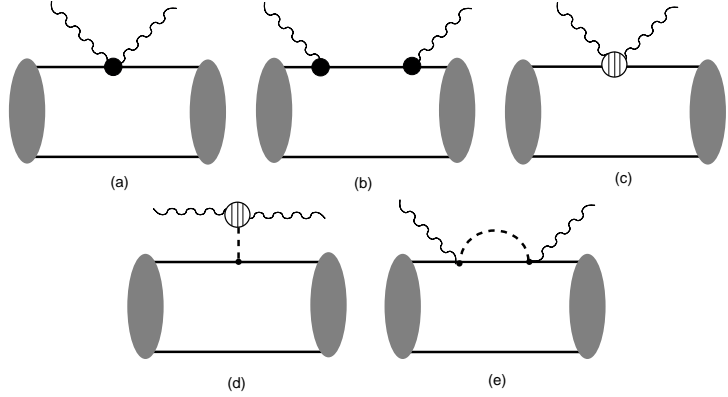}}
   \caption{\label{fig1} Characteristic
   one-body interactions that contribute to Compton scattering on the
   deuteron at order $Q^2$ (a) and at order $Q^3$ (b-e) (in the
   Coulomb gauge). Crossed graphs are not shown and not all loop
   topologies are shown.  The sliced blobs are vertex insertions from
   ${\cal L}^{(1)}$. The sliced and diced blobs are vertex insertions
   from ${\cal L}^{(2)}$.}
\end{figure}

The leading contributions to the irreducible $\gamma NN \rightarrow
\gamma NN$ kernel $K_{\gamma \gamma}$ appear at $\nu
=-1$---$O(Q^2)$---from a tree-level diagram ($L=0$) with two
separately connected pieces ($C=2$) and a vertex which is the
two-photon seagull of $\Delta _i =1$. This vertex arises from minimal
coupling in the kinetic nucleon terms in ${\cal L}^{(1)}$ of
Eq.~(\ref{L1})---see Fig.~\ref{fig1}(a). It therefore represents
Compton scattering on the $NN$ system where the neutron
is a spectator while the photon scatters from the proton
via the Thomson amplitude (\ref{eq:OQ2}).

\begin{figure}[t,h,b,p]
  \epsfysize=6.5cm
   \centerline{\epsffile{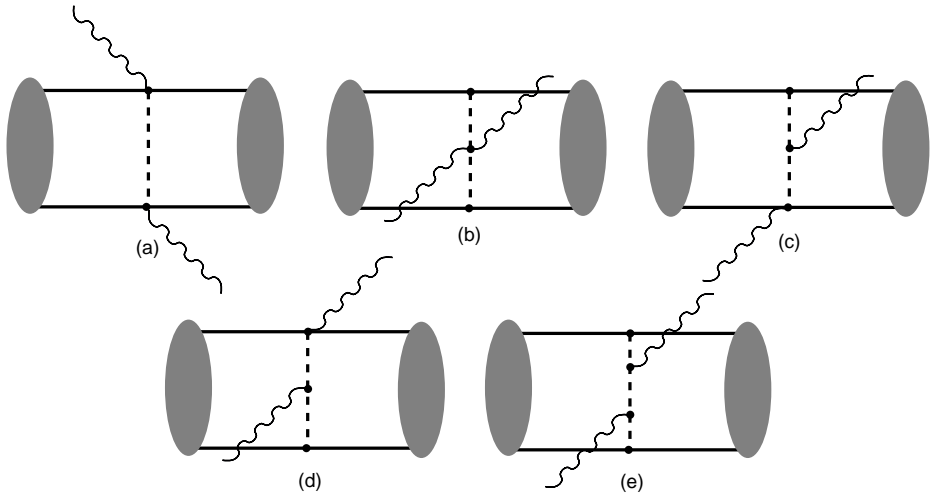}}
   \caption{\label{fig-oq3diags} Two-body
   interactions which contribute to Compton scattering on the deuteron
   at order $Q^3$ ($\nu=0$) in Coulomb gauge. Graphs which differ only
   by interchange of the nucleons are not shown.}
\end{figure}

At orders $\nu \ge 0$ there are two classes of corrections:
corrections to the one-nucleon amplitude where the second nucleon is
just a spectator, and corrections that involve both nucleons.  Shown
in Fig.~\ref{fig1}(b-e) are the one-nucleon corrections at $\nu = 0$,
while Fig.~\ref{fig-oq3diags} displays two-nucleon corrections of the
same order. Note that the one-nucleon-loop and two-nucleon diagrams
involve only $\Delta _i =0$ interactions in diagrams with $L=1$ or
$C=1$.

\begin{figure}[thbp]
   \epsfysize=4.0cm
   \centerline{\epsffile{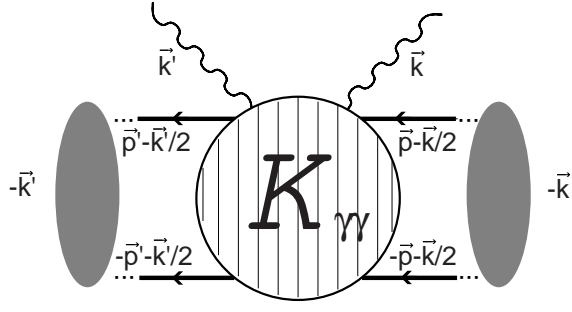}}
   \caption{\label{fig-kinematics}Kinematics
   for the $\gamma$d process in the $\gamma$d center-of-mass
   frame. The sliced circle represents $K_{\gamma \gamma}$, the
   irreducible $\gamma NN \rightarrow \gamma NN$ kernel.}
\end{figure}

At this, or indeed any, order the irreducible $\gamma NN
\rightarrow \gamma NN$ kernel can be separated into 
one- and two-body pieces. Doing this in the $\gamma NN$ c.m. frame,
with the kinematics shown in Fig.~\ref{fig-kinematics}
we obtain:
\begin{equation}
K_{\gamma \gamma}^{\gamma NN
c.m.}(\vpeeprime,\vkayprime; \vpee,\vkay)= T^{\gamma d \, c.m.}_{\gamma
N}(\vkayprime;\vkay) \, \, \delta^{(3)}\left(p' - p
- \smfrac{1}{2}q\right) + T^{2 N}_{\gamma NN}
(\vpeeprime,\vkayprime; \vpee,\vkay);
\label{eq:decomp}
\end{equation}
with $\vec{q} \equiv \vec{k} - \vkayprime$ the momentum transfer of
the Compton scattering process.  The one-nucleon piece $T^{\gamma d \,
c.m.}_{\gamma N}(\vkayprime;\vkay)$ is related to the $\gamma N$
amplitude discussed in Sec.~\ref{sec-prottheory} by a boost from the
Breit frame to the $\gamma$d c.m. frame. Details of the boosting
procedure are given in the Appendices.

Meanwhile $T_{\gamma NN}^{2 N}$ was computed 
to third order in small quantities in Ref.~\cite{therestofus}.
At fourth order the difference of initial and final nucleon kinetic
energies should be included in the pion propagators appearing in
Fig.~\ref{fig-oq3diags}. The result is then:
\begin{equation}
T_{\gamma NN,\,O(Q^3)}^{2N}\;=\;-\frac{{e^2}{\ga ^2}}{2
f_\pi^2} \; ({\vec\tau}^{\; 1} \cdot{\vec\tau}^{\;
2}-\tau^{1}_{z}\tau^{2}_{z}) \; ( t^{(a)}+ t^{(b)}+ t^{(c)}+
t^{(d)}+ t^{(e)}) \label{total}
\end{equation}
with:
\begin{eqnarray}
{t^{(a)}}&= &
\frac{{\veps\cdot\vsigone}\;{\vepsprime\cdot\vsigtwo}}
{2\lbrack \tilde{\omega}^2-{m_\pi^2}-
(\vpee -\vpeeprime + \vkay_+)^2 \rbrack}
+ (1\;\leftrightarrow\; 2)
\label{ta}\nonumber\\
{t^{(b)}}&= &
\frac{{\veps\cdot\vepsprime}\;
\vsigone\cdot ( \vpee -\vpeeprime -\smfrac{1}{2}\vec{q}^{\,} )
        \vsigtwo\cdot ( \vpee -\vpeeprime + \smfrac{1}{2} \vec{q}^{\,})}
{2\lbrack (\vpee -\vpeeprime -\smfrac{1}{2}\vec{q}^{\,})^2
+{m_\pi^2} \rbrack
\lbrack (\vpee -\vpeeprime +\smfrac{1}{2}\vec{q}^{\,})^2
+m_\pi^2\rbrack}
+ (1\;\leftrightarrow\; 2)
\label{tb}\nonumber\\
{t^{(c)}}&= &
-\frac{{\vepsprime\cdot (\vpee -\vpeeprime +{\frac{1}{2}\vkay})}\;
          \vsigone\cdot\veps\;
 \vsigtwo\cdot ( \vpee -\vpeeprime +\smfrac{1}{2}\vec{q}^{\,})}
{\lbrack \tilde{\omega}^2-m_\pi^2- (\vpee -\vpeeprime
+\vec{k}_+)^2 \rbrack
\lbrack (\vpee -\vpeeprime + \frac{1}{2}\vec{q}^{\,})^2
+m_\pi^2 \rbrack}
+ (1\;\leftrightarrow\; 2)
\nonumber\\
{t^{(d)}}&= &
-\frac{{\veps\cdot (\vpee -\vpeeprime +{\smfrac{1}{2}\vkayprime} )}\;
\vsigone\cdot ( \vpee -\vpeeprime -{\smfrac{1}{2}\vec{q}^{\,})} )
              \vsigtwo\cdot\vepsprime }
{\lbrack \tilde{\omega}^2-{m}_\pi^2- (\vpee -\vpeeprime +
\vec{k}_+)^2 \rbrack
\lbrack (\vpee -\vpeeprime -\smfrac{1}{2}\vec{q}^{\,})^2
+m_\pi^2 \rbrack}+ (1\;\leftrightarrow\; 2)\nonumber\\
{t^{(e)}}&= &
\frac{
\vsigone\cdot ( \vpee -\vpeeprime -\smfrac{1}{2}\vec{q}^{\,})\;
\vsigtwo\cdot ( \vpee -\vpeeprime +\smfrac{1}{2}\vec{q}^{\,})}
{\lbrack (\vpee -\vpeeprime -\smfrac{1}{2}\vec{q}^{\,})^2
+{m_\pi^2} \rbrack
\lbrack (\vpee -\vpeeprime +\smfrac{1}{2}\vec{q}^{\,})^2
+{m_\pi^2} \rbrack}\nonumber \\
&& \qquad \qquad \qquad \times \frac{2 \; \veps\cdot (\vpee -\vpeeprime +{\frac{1}{2}\vkayprime})\;
\vepsprime\cdot (\vpee -\vpeeprime +{\frac{1}{2}\vkay})}
{\lbrack \tilde{\omega}^2
-{m_\pi^2}- (\vpee -\vpeeprime +
\vec{k}_+)^2 \rbrack}
\;+\; (1\;\leftrightarrow\; 2),
\label{te}
\end{eqnarray}
where $\vec{k}_+ \equiv \smfrac{1}{2}(\vkay + \vkayprime)$ is the 
average of incoming and outgoing photon three-momenta, the notation
$(1 \leftrightarrow 2)$ indicates the inclusion of a second
term in which $\vpee \rightarrow -\vpee$, $\vpeeprime \rightarrow -
\vpeeprime$, $\vsigone \leftrightarrow \vsigtwo$,
and:
\begin{equation}
\tilde{\omega}^2 \equiv \omega^2 + \omega \, \frac{\vpeeprime \cdot \vkayprime 
- \vpee \cdot \vkay}{M}.\label{eq:omegatilde}
\end{equation}

In computing the ``retardation'' effects which modify modify
$\omega\to\tilde\omega$ we use the fact that---because we are
considering an elastic process---the two-nucleon initial and final
states are equally distant from being on-shell. Thus the results
(\ref{te}), (\ref{eq:omegatilde}) were derived using
$|\vpee^{\,}|^2=|\vpeeprime|^2$, which simplifies the expression for
$\tilde{\omega}$.  Note that for virtual momenta $\vpee$ and
$\vpeeprime$ of order $M$ the retardation effects included here will
have unphysical consequences, and thus Eq.~(\ref{eq:omegatilde}) is
only valid if $|\vpee^{\,}|,|\vpeeprime| \ll M$.

Upon including these retardation effects in our calculation we found
them to be numerically very small. Hence in the computations of the
$\gamma$d differential cross section which we report on below they
were not considered.

Since neither the expression for $T^{2N}_{\gamma NN, O(Q^3)}$ nor the
$\gamma N$ $O(Q^3)$ amplitude contain any free parameters, $\chi$PT
makes predictions for $\gamma$d scattering at this order. These
predictions were generated in Ref.~\cite{therestofus} and compared to
the data of Ref.~\cite{lucas}, with encouraging results, especially at
$E_{\rm lab}=69$ MeV. The $O(Q^3)$ predictions (made on the basis of
Ref.~\cite{therestofus}) for the Lund experiments at $\omega \approx
55$ MeV and $\omega \approx 66$ MeV turned out to be in excellent
agreement with that data too.  (See curves below.)

At $O(Q^4)$ there are many one-nucleon diagrams in the $\gamma NN$
kernel, which are easily obtained from those in Figs.~\ref{protborn}
and \ref{protloop}. These were computed in the Breit frame in
Ref.~\cite{judith}. The extra pieces of the $\gamma N$ amplitude which
are obtained in the $\gamma$d center-of-mass frame as compared to the
Breit-frame result are discussed in Appendices
\ref{ap-oq3prot}--\ref{ap-gammadframe}.

\begin{figure}[tbph]
  \begin{center} \mbox{\epsfig{file=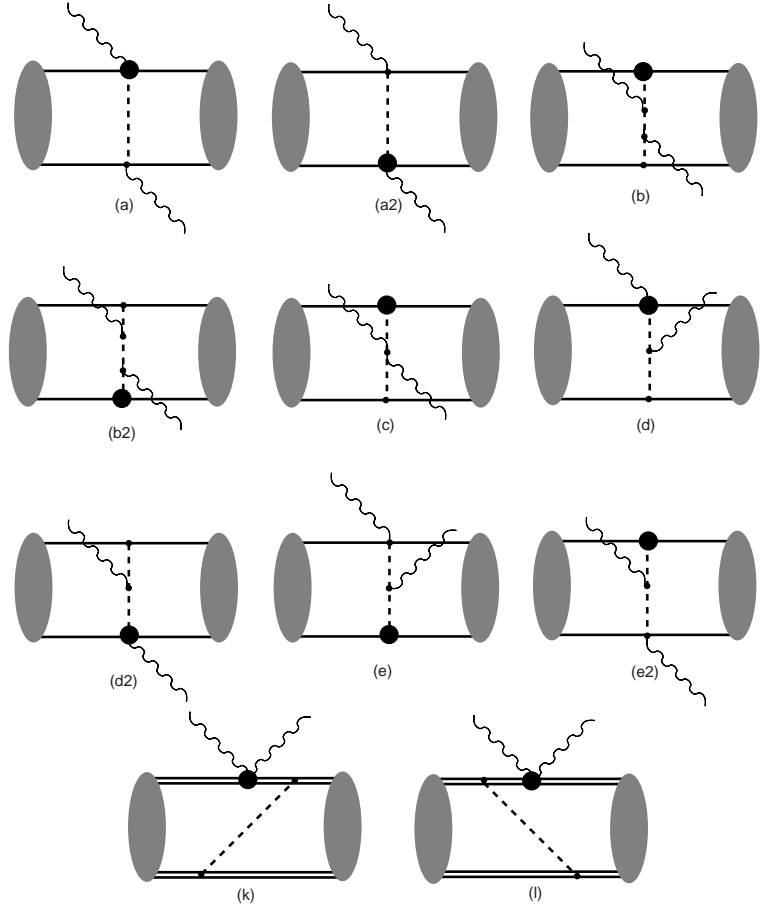,width=10truecm,angle=0}}
  \end{center}
   \caption{\label{fig-zerographs} Some of the two-body graphs that
are nominally $O(Q^4)$ for $\gamma$d scattering in Coulomb
gauge.  Ultimately all of the graphs shown here are either identically
zero or suppressed to $O(Q^5)$ by kinematic effects, as explained in
the text.
Non-zero two-body contributions to $\gamma$d scattering
at $O(Q^4)$ are shown in Fig.~\ref{fig-oq4diags}.\hfill.}
\end{figure}

\begin{figure}[tbph]
  \begin{center} \mbox{\epsfig{file=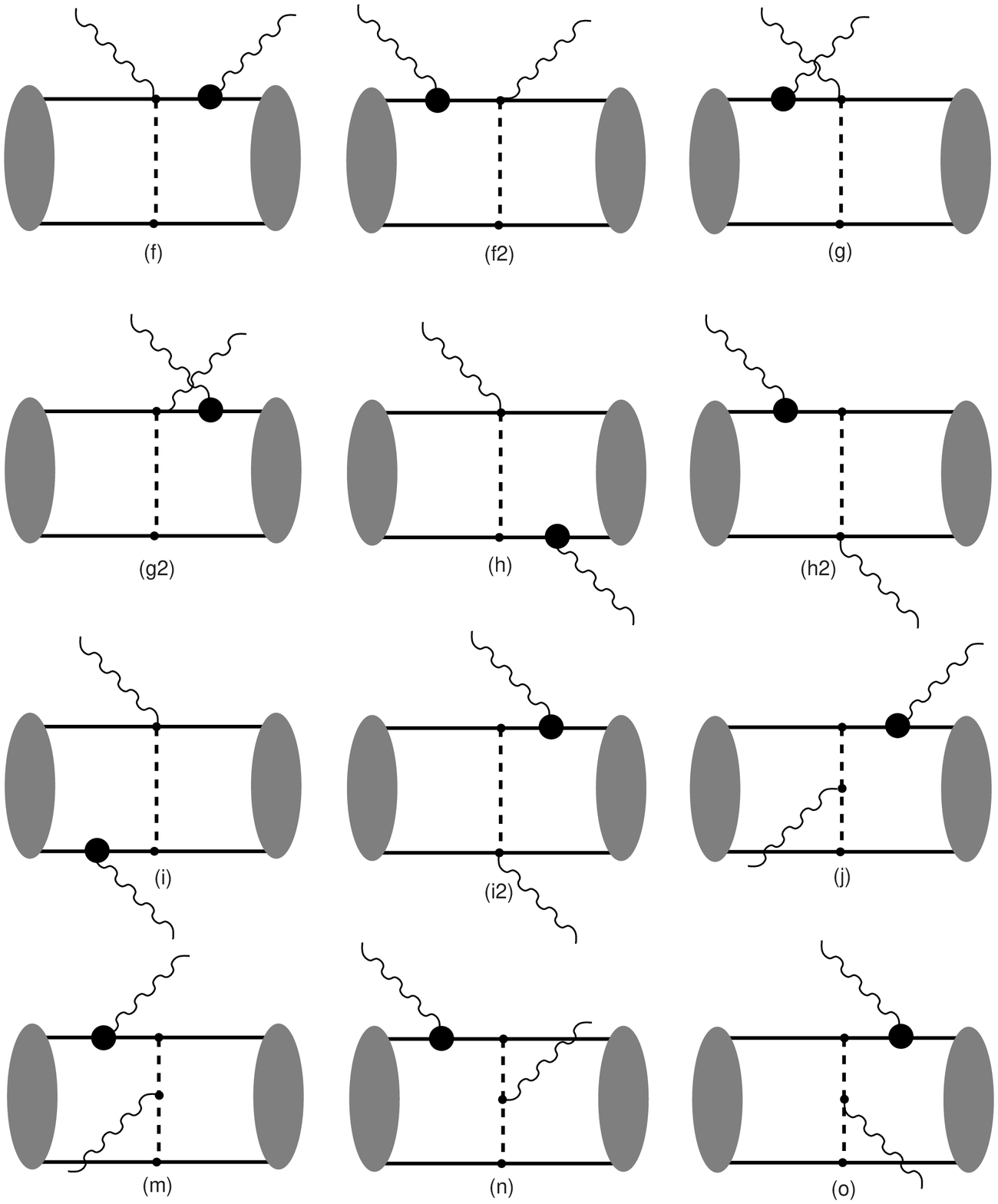,width=11truecm,angle=0}}
  \end{center}
   \caption{\label{fig-oq4diags}
The two-body diagrams contributing to $\gamma$d scattering
in Coulomb gauge at $O(Q^4)$.}
\end{figure}

Two-nucleon diagrams at $O(Q^4)$ in Coulomb gauge are depicted in
Figs.~\ref{fig-zerographs} and \ref{fig-oq4diags}.  These two-nucleon
$\nu=1$ diagrams have $C=1$ with one insertion from ${\cal L}^{(1)}$.

Although they are nominally of $O(Q^4)$, all of the graphs in
Fig.~\ref{fig-zerographs} give zero contribution to the elastic
$\gamma$d amplitude at this order.  Diagrams (a), (a2), (d), and (d2)
do not affect this process because $K_{\gamma \gamma}$ is
of isovector character. Meanwhile graphs (b), (b2), (c), (e), and (e2)
all involve the $\pi NN$ vertex from ${\cal L}^{(1)}$. With the choice
$v=(1,\vec{0})$ the Feynman rule for this vertex reads:
\begin{equation}
\frac{g_A}{4 M f_\pi} \, \vec{\sigma} \cdot (\vec{p} + \vpeeprime) \, 
q_0 \, \tau^a,
\end{equation}
where $\vec{p}$ ($\vpeeprime$) is the nucleon momentum before (after)
the emission of the pion, and $q$ is the emitted pion's
four-momentum. In the context of the $\gamma NN$ kernel being
discussed here the emitted pion is a ``potential'' pion~\cite{Ka98},
and as such has $q_0 \sim {\vpee \,}^2/M$. Consequently in the
kinematics we are considering all five of these graphs are suppressed
to $O(Q^5)$. Lastly, graphs (k) and (l) represent time-orderings where
the photon interacts with a single nucleon while the pion is
``in-flight''. If, as is the case here, an instantaneous potential is
employed to generate the deuteron wave function then consistency
between the potential $V$ and the $\gamma NN \rightarrow \gamma NN$
kernel requires that graphs (k) and (l) not be included in $K_{\gamma
\gamma}$.

The non-zero two-body graphs at $O(Q^4)$ are depicted in
Fig.~\ref{fig-oq4diags}.  The ($\gamma$d c.m.~frame)
expressions for these pieces of the irreducible $\gamma NN
\rightarrow \gamma NN$ kernel sum to:
\begin{equation}
T_{\gamma NN,\,O(Q^4)}^{2N}\;=\;\;\frac{{e^2}{\ga ^2}}{4
f_\pi^2} \; \frac{1}{M \omega}\;({\vec\tau}^{\; 1}
\cdot{\vec\tau}^{\; 2}-\tau^{1}_{z}\tau^{2}_{z}) \; ( t^{(f+g)}+
t^{(f_2+g_2)}+ t^{(h+i+h_2+i_2)}+ t^{(j+m)}+ t^{(n+o)}),
\label{q4tb}
\end{equation}
with:
\begin{eqnarray}
t^{(f+g)}&=&-\frac{1}{2[\qtwoline^2+m_\pi^2]}\{\vepsprime
\cdot (\vpee + \vpeeprime - \vec{k}_+)
\nonumber\\
&& \times [\vsigone\cdot\qtwoline\vsigtwo\cdot\veps+\vsigtwo\cdot\qtwoline
\vsigone\cdot\veps]\nonumber\\
&&+ i (1+\kappa_{v})\veps\cdot(\vepsprime\times\vkayprime)
(\vsigone+\vsigtwo)\cdot\qtwoline\}\nonumber\\
t^{(f_2+g_2)}&=&\frac{1}{2[\qtwoline^2+m_\pi^2]}\{\veps\cdot
(\vpee + \vpeeprime - \vec{k}_+)
\nonumber\\&&\times[\vsigone\cdot\qtwoline\vsigtwo\cdot\vepsprime+\vsigtwo\cdot
\qtwoline\vsigone\cdot\vepsprime] \nonumber
\\&&-i(1+\kappa_{v})\vepsprime
\cdot(\veps\times\vkay)(\vsigone+\vsigtwo)\cdot\qtwoline\}\nonumber\\
t^{(h+i+h_2+i_2)}&=&\frac{1}{2[\omega^{2}-m_\pi^2-\qnonline^2]}
\{\vepsprime\cdot (\vpee + \vpeeprime + \vec{k}_+)
\nonumber\\
&&\times[\vsigone\cdot\qnonline\vsigtwo\cdot\veps+\vsigtwo\cdot
\qnonline\vsigone\cdot\veps] \nonumber\\
&&-i(1+\kappa_{v})\qnonline
\cdot(\vepsprime\times\vkayprime)\, (\vsigone+\vsigtwo)
\cdot\veps\nonumber\\&&
-\veps\cdot(\vpee + \vpeeprime - \vec{k}_+)
[\vsigone\cdot\qnonline\vsigtwo\cdot
\vepsprime\nonumber\\
&&\qquad \qquad + \vsigtwo\cdot\qnonline\vsigone\cdot\vepsprime]\nonumber\\&&+ i(1+
\kappa_{v})\qnonline\cdot(\veps\times\vkay)(\vsigone+\vsigtwo)
\cdot\vepsprime\}\nonumber\\
t^{(j+m)}&=&-\frac{1}{[\omega^{2}-m_\pi^2-\qnonline^2]}\frac{1}
{[m_\pi^2+\qoneline^2]}\nonumber\\
&&\times \veps
\cdot(\vpee-\vpeeprime+ \vec{k}_+)\{\vepsprime
\cdot(\vpee + \vpeeprime + \vec{k}_+)\nonumber\\
&& \times [\vsigone\cdot\qnonline\vsigtwo\cdot\qoneline\nonumber\\
&& \qquad \qquad +  \vsigtwo\cdot\qnonline\vsigone\cdot\qoneline]\nonumber\\
&&-i (1+\kappa_{v})(\vepsprime\times\vkayprime)\cdot\qnonline
(\vsigone+\vsigtwo)\cdot\qoneline\}\nonumber\\
t^{(n+o)}&=&\frac{1}{[\omega^{2}-m_\pi^2-\qnonline^2]}
\frac{1}{[m_\pi^2+\qtwoline^2]}\nonumber\\&&\times\vepsprime
\cdot(\vpee-\vpeeprime+\vec{k}_+)\{\veps\cdot(\vpee + \vpeeprime - \vec{k}_+)
\nonumber\\&&\times[\vsigone\cdot\qnonline\vsigtwo\cdot\qtwoline\nonumber
\nonumber\\
&&\qquad \qquad 
+\vsigtwo\cdot \qnonline\vsigone\cdot\qtwoline]\nonumber\\&&-i(1+\kappa_{v})\qnonline(\veps\times\vkay) (\vsigone+\vsigtwo)\cdot\qtwoline\}.
\end{eqnarray}
(In this case the contribution from diagrams related to those in
Fig.~\ref{fig-oq4diags} by interchange of the two nucleons has been
explicitly included in the expressions.)

The vertices from the next-to-leading-order $\pi N$ Lagrangian which
appear in these non-zero two-body diagrams all involve E1 and M1
couplings of the photon to the nucleons. Thus the only parameters
appearing are the proton charge and the proton and neutron anomalous
magnetic moments. {\it There are no free parameters in the two-body
$O(Q^4)$ contribution.} This is good, as these mechanisms are of the
same order as the counterterms $\delta \alpha_N$ and $\delta \beta_N$
which we are trying to fit. Extracting the polarizabilities from
deuteron Compton data is much more straightforward if the $O(Q^4)$
two-body currents can be expressed in terms of known parameters.

To calculate the amplitude for $\gamma$d scattering, $T_{\gamma d}$, we
first sandwich $K_{\gamma \gamma}$ between deuteron wave functions and
use the decomposition of Eq.~(\ref{eq:decomp}). This yields (in the
$\gamma$d center-of-mass frame):
\begin{eqnarray}
T^{\gamma d}_{M' \lambda' M \lambda}(\vkayprime,\vkay)&=& \int
\frac{d^3p}{(2 \pi)^3} \, \, \psi_{M'}^*\left( \vpee + \smfrac{1}{2}
\vec{q}^{\,}\right) \, \, T^{\gamma d \, \, c.m.}_{\gamma N_{\lambda'
\lambda}}(\vkayprime,\vkay) \, \, \psi_M(\vpee)\nonumber\\
&+& \int \frac{d^3p \, \, d^3p'}{(2 \pi)^6} \, \, \psi_{M'}^*(\vpeeprime) \, \,
T^{2N}_{\gamma NN_{\lambda' \lambda}}(\vkayprime,\vkay) \, \, \psi_M(\vpee)
\label{eq:gammad}
\end{eqnarray}
where $M$ ($M'$) is the initial (final) deuteron spin state, and
$\lambda$ ($\lambda'$) is the initial (final) photon polarization
state, and $\vkay$ ($\vkayprime$) the initial (final) photon
three-momentum, which are constrained to
$|\vkay|=|\vkayprime|=\omega$. 

The laboratory differential cross section can then be evaluated directly:
\begin{equation}
  \frac{d \sigma}{d \Omega_L}=\frac{1}{16 \pi^2}
  \left(\frac{E_\gamma'}{E_\gamma}\right)^2 \frac{1}{6} \sum_{M'
    \lambda' M \lambda} |T^{\gamma d}_{M' \lambda' M \lambda}|^2,
\label{eq:labdcs}
\end{equation}
where $E_\gamma$ is the initial photon energy in the laboratory frame,
and is related to $\omega$, the photon energy in the $\gamma$d
center-of-mass frame, via:
\begin{equation}
\omega=\frac{E_\gamma}{\sqrt{1 + 2 E_\gamma/M_d}};
\end{equation}
and $E_\gamma'$ is the final photon energy in the laboratory frame:
\begin{equation}
E_\gamma'=\frac{E_\gamma M_d}{M_d + E_\gamma (1 - \cos \theta_L)}.
\end{equation}
To evaluate the center-of-mass-frame differential cross section
we employ:
\begin{equation}
\frac{d \sigma}{d \Omega_{cm}}=
\frac{M_d}{16 \pi^2 (M_d + 2 E_\gamma)} 
\frac{1}{6} \sum_{M'
    \lambda' M \lambda} |T^{\gamma d}_{M' \lambda' M \lambda}|^2.
\label{eq:cmdcs}
\end{equation}

\section{Compton scattering on the deuteron: results}
\label{sec-results}

In this section we present our results for the cross section for
Compton scattering on the deuteron including the one-nucleon and
two-nucleon mechanisms described above.  Our calculation therefore
represents the full $\nu =1$, or $O(Q^4)$, $\chi$PT result for Compton
scattering on the deuteron.  We present these $O(Q^4)$ results,
examine how far they depend on the choice of deuteron wave function,
and discuss our calculation's breakdown as the photon energy is
reduced to energies comparable to the nuclear binding scale,
$m_\pi^2/M$. We then present results for different fits to the
database of recent $\gamma$d experiments. Because the deuteron is an
isoscalar, the only free parameters are the isoscalar combinations
$\delta \alpha_N \equiv (\delta\alpha_p + \delta\alpha_n)/2$ and
$\delta \beta_N \equiv (\delta\beta_p + \delta\beta_n)/2$. In fact, of
all the additional terms which appear in the $\gamma$d calculation
when we go from $O(Q^3)$ to $O (Q^4)$, only these LECs affect the
cross section significantly. We conclude the section by comparing our
results with numbers for neutron polarizabilities obtained via other
techniques.

\begin{figure}[t!]
\vspace{0.2in}
\epsfxsize=10.0cm
   \centerline{\epsffile{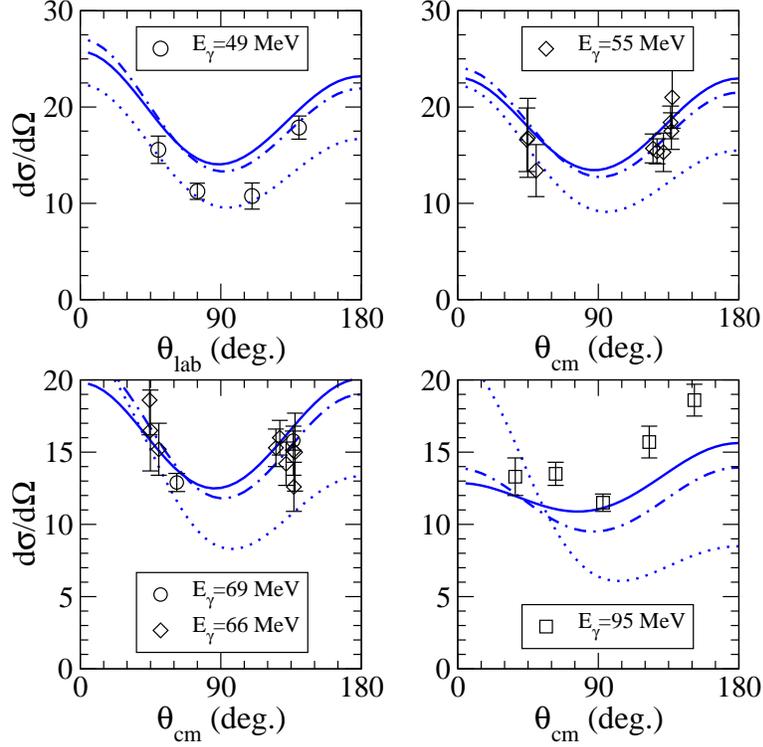}}
\caption {\label{fig-fitI} Results of the $O(Q^2)$,
$O(Q^3)$ and $O(Q^4)$ calculations of (lab and c.m. as appropriate)
differential cross sections for Compton scattering on deuterium at
four different lab photon energies: 49, 55, 67, and 94.5 MeV. The data
are from Illinois~\cite{lucas} (circles), Lund~\cite{lund} (diamonds)
and SAL (squares)~\cite{SAL}. The solid line is the $O(Q^4)$
calculation with $\alpha_N=13.6 \times 10^{-4}~{\rm fm}^3$,
$\beta_N=0.1 \times 10^{-4}~{\rm fm}^3$ (Fit I of
Table~\ref{table-results}).  The dot-dashed line is the (parameter-free)
$O(Q^3)$ calculation, and the dotted line is the result of the
leading-order [$O(Q^2)$] calculation.}
\end{figure}

\subsection{Results at $O(Q^4)$ and convergence of \cpt}

Our results for Compton scattering on deuterium at $O(Q^4)$ are
presented in Fig.~\ref{fig-fitI}, where the solid line
represents the calculations at photon energies of 49 MeV (lab), and
55, 66, and 95 MeV (c.m.). Also shown are the results of lower-order
calculations at the same energies, as well as data from
Refs.~\cite{lucas,SAL,lund}. The wave function employed was the NLO
$\chi$PT wave function of Ref. \cite{Ep99}, with the cutoff $\Lambda$
chosen to be $600$ MeV. As emphasized above, the only free parameters
in this calculation are the isoscalar polarizabilities $\alpha_N$ and
$\beta_N$. The $O(Q^4)$ plots in the figure were generated with the
values for $\alpha_N$ and $\beta_N$ obtained in Fit I of
Table~\ref{table-results} below.  However, before discussing this, and
our other fits, in detail, we want to use these results to address a
number of theoretical issues associated with our calculation.

One of the strengths of effective field theory is that it establishes
a perturbative expansion for S-matrix elements---see
Eq.~(\ref{pionfulT}). Having employed an EFT to compute $\gamma$d
scattering it behooves us to ask whether the perturbation expansion is
behaving as expected. To this end in Fig.~\ref{fig-fitI} we
also plot the $\gamma$d cross section at leading ($O(Q^2)$, LO, dotted
line), and next-to-leading ($O(Q^3)$, NLO, dot-dashed line) order. The
correction from LO to NLO is sizable, but is consistent with an
expansion parameter $\smfrac{\omega}{M_\Delta - M}$.  The
effect in going from NLO to N$^2$LO is surprisingly small, perhaps
because two-body effects that enter are controlled by, at worst,
$\omega/M$, which is significantly smaller than the nominal expansion
parameter $\omega/(M_\Delta - M)$.  Generically, the largest effect at
N$^2$LO comes from the shifts of the polarizabilities from their
$O(Q^3)$ values. Since this effect goes as $\omega^2$ the shift in the
$O(Q^4)$ result compared to that at $O(Q^3)$ is much more noticeable
at 95 MeV than at any of the lower energies.

All of this is encouraging for our attempts to extract $\alpha_N$ and
$\beta_N$ from the coherent Compton cross section. On the other hand,
a skeptical view of the results in Fig.~\ref{fig-fitI} would be
that another order must be calculated in order to ensure
convergence. We will return to this point below, but we note that the
as-yet-uncomputed fifth-order $\chi$PT result for the $\gamma N$ amplitude
is a key element in any such calculation. The $O(Q^5)$ result for
$\gamma$d scattering therefore requires a two-loop computation in
the single-nucleon sector.

\subsection{Wave-function dependence}

We have computed the $\gamma NN \rightarrow \gamma NN$ kernel 
to two orders beyond leading order.  For computation of the matrix element that
enters observables the NLO wave functions of Ref. \cite{Ep99}
are therefore a consistent choice since they too include effects of
$O(Q^2)$ beyond leading.  (They are ``next-to-leading order'' for $NN$
scattering, since the $O(Q)$ corrections to $NN$ vanish in the
parity-conserving part of the $NN$ potential.)

Ref. \cite{Ep99} employed a cutoff in its NLO calculation: a
cutoff that was varied between 500 and 600 MeV. The results of our
$\gamma$d computation do not depend markedly on this
cutoff. However, our results are rather sensitive to which $NN$
potential is chosen in order to generate the deuteron wave
function. In particular, the Nijm93 $NN$ potential-model~\cite{Nijm93}
gives a deuteron wave function which, when used in
Eq.~(\ref{eq:gammad}), results in cross sections somewhat larger than those
found with the wave function of Ref. \cite{Ep99}. These two results
are extremal, in the sense that other wave functions (e.g. the Bonn OBEPQ
wave function~\cite{Bonnrep}, and the simple one-pion-exchange plus
square-well wave functions of Ref.~\cite{PC99}) generate
cross sections which fall in between them.

\begin{figure}[t!]
\vspace{0.2in}
\epsfxsize=10.0cm
   \centerline{\epsffile{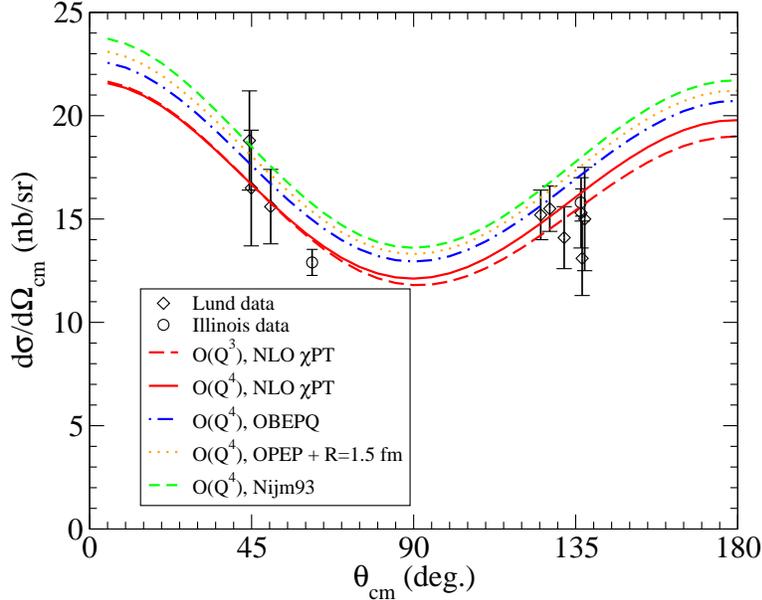}}
\caption {\label{fig-wfdep} $O(Q^4)$ calculation of c.m.  differential
cross section for $\gamma$d scattering with $\omega=66$ MeV. Data from
Illinois and Lund, as in Fig.~\ref{fig-fitI}.  The dashed
line is the (parameter-free) $O(Q^3)$ calculation with the NLO
$\chi$PT wave function. In this curve the polarizabilities have their
$O(Q^3)$ values. All other curves were obtained using the $O(Q^4)$
$\gamma NN \rightarrow \gamma NN$ kernel, with $\alpha_N=13.6 \times
10^{-4}~{\rm fm}^3$ and $\beta_N=0.1 \times 10^{-4}~{\rm fm}^3$. The
solid red line is the result with the NLO $\chi$PT wave function of
Ref.~\cite{Ep99}, and the short-dashed green, dot-dashed blue and
dotted orange lines use, respectively, the Nijm93, the Bonn OBEPQ, and
$R=1.5$ fm + OPEP~\cite{PC99} wave functions. In all cases the cutoff
$\bar{\Lambda}$ is chosen to be 600 MeV.}
\end{figure}

In Fig.~\ref{fig-wfdep} we present a selection of these results for
the case $\omega=66$ MeV. Different wave functions yield cross section
predictions which vary by about 10-20\%, with greater variation at
backward angles. The variation with choice of wave function decreases
slightly with energy. This wave-function dependence comes almost
entirely from the matrix elements of the two-body operators.  The
impulse-approximation piece of the cross section probes deuteron
structure at momentum transfers of 200 MeV or less, and all of these
wave functions give very similar results in that domain.

Our $\gamma NN \rightarrow \gamma NN$ kernel is valid only at low
momenta. Thus, in all results presented here we have introduced a cutoff
$\bar{\Lambda}$ on the momenta $p$ and $p'$ in the six-dimensional
integral of Eq.~(\ref{eq:gammad}). The cutoff function employed is:
\begin{equation}
f(k)=\exp(-k^4/\bar{\Lambda}^4),
\label{eq:cutoff}
\end{equation}
in imitation of the wave-function calculation of Ref.~\cite{Ep99}. For
all our results we chose $\bar{\Lambda}=600$ MeV, so as to be
consistent with the $\Lambda=600$ MeV wave function of
Ref.~\cite{Ep99}.  Provided $\bar{\Lambda} \geq 600$ MeV, inserting the
cutoff function in the matrix element evaluation changes the
results with the NLO $\chi$PT wave function by less than $2\%$. It
does, however, eliminate some high-momentum strength in the matrix
elements evaluated with the Nijm93 wave function, thereby reducing the
cross section. Not including the cutoff function (\ref{eq:cutoff})
increases the spread of cross sections in Fig.~\ref{fig-wfdep} to 30\%
or more at 66 MeV.  Furthermore, without such a cutoff, wave functions
from potentials with deeply-bound states, such as the NNLO wave
function of Ref.~\cite{Ep99}, generate cross sections 10--100 times
larger than those displayed here. With the cutoff in place, and
$\bar{\Lambda}=600$ MeV, the NNLO wave function of Ref.~\cite{Ep99}
gives a result that falls within the band defined by the NLO $\chi$PT
wave function and the Nijm93 wave function.

\subsection{Modifying the power counting for $\omega \sim \frac{m_\pi^2}{M}$}

\label{sec-resum}

\begin{figure}[t!]
  \begin{center} \mbox{\epsfig{file=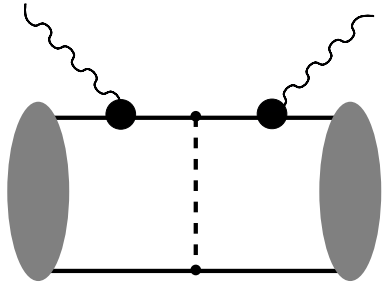,width=5truecm,angle=0}}
  \end{center}
\caption{\label{fig-example} One Feynman diagram which is of a higher
order in the chiral expansion than we consider here, but must be
included in the calculation of the $\gamma$d amplitude if $T_{\gamma
d}$ is to be exactly gauge invariant. The shaded blobs represent
deuteron vertex functions. The small dots are vertices from the
leading-order $\pi N$ Lagrangian, ${\cal L}^{(0)}$, while the larger
ones are vertices from ${\cal L}^{(1)}$.}
\end{figure}

No matter which wave function is chosen our $O(Q^3)$ and $O(Q^4)$
calculations overestimate the 49 MeV cross section, as measured at
Illinois by Lucas~\cite{lucas}. This would seem to be connected to the
fact that at photon energies $\omega \sim \smfrac{m_\pi^2}{M}$ the
power counting employed here breaks down and the contribution from
two-nucleon-intermediate states must be included in full in the
calculation. As things stand diagrams such as Fig.~\ref{fig-example}
only occur in the chiral expansion of the $\gamma NN$ kernel at
$O(Q^5)$ and beyond, and without the inclusion of graphs like this one
$T_{\gamma d}$ is not gauge invariant. Thus the power counting we have
used so far for $\gamma$d does not recover the Thomson-limit amplitude
for Compton scattering on deuterium as $\omega \rightarrow 0$.

These difficulties stand in contrast to the EFT($\not \! \pi$) calculations of
Refs.~\cite{griess}, which recover the $\gamma$d Thomson limit
and also reproduce the 49 MeV Illionis data.  In
the nomenclature of Refs.~\cite{griess,Ch98} our calculation of
$\gamma$d scattering is valid in ``Regime II'': the kinematic domain
where $\omega$ is of order the deuteron binding momentum: $\omega \sim
\gamma \equiv \sqrt{MB}$. The Thomson limit for $\gamma$d scattering
emerges in a different regime: ``Regime I'' , which corresponds to
photon energies of order the deuteron binding energy, i.e. $\omega
\sim B$.

\begin{figure}[t!]
\vspace{0.2in}
\epsfxsize=12.0cm
   \centerline{\epsffile{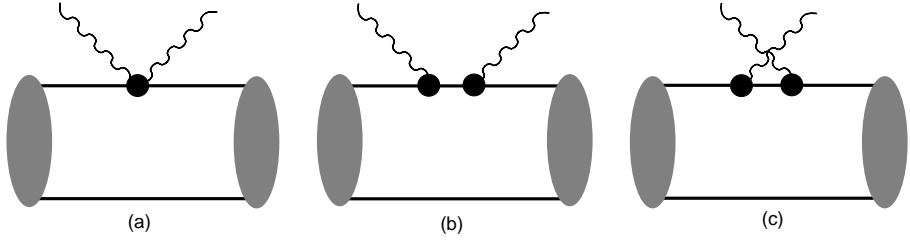}}
\caption {\label{fig-bddiags} Leading diagrams in Regime I
for $\gamma$d scattering in EFT($\not \! \pi$).}
\end{figure}

We now repeat the argument of Ref.~\cite{Ch98} in order to shed some
light on how the $\gamma$d Thomson limit emerges in Regime I. There
the leading diagrams for Compton scattering on deuterium are shown in
Fig.~\ref{fig-bddiags}. The expressions for the operators to be
sandwiched between deuteron wave functions can be written (in the
$\gamma$d c.m. frame):
\begin{eqnarray}
\hat{O}^{(a)}&=&-\frac{e^2}{M} \vepsprime \cdot \veps;\nonumber\\
\hat{O}^{(b)}&=&-\frac{e^2}{M^2} \vepsprime \cdot \vpee
\frac{1}{\omega + \smfrac{\omega^2}{2 M_d} - B - 
\frac{\vpee^{\, 2}}{M}}\veps \cdot \vpee\nonumber\\
\hat{O}^{(c)}&=&-\frac{e^2}{M^2} 
\veps \cdot \vpee \frac{1}{\smfrac{\omega^2}{2M_d} - B -
\omega - \frac{(\vpee - 2 \vec{k}_+)^2}{2 M}
- \frac{{\vpee \,}^2}{2 M}} \vepsprime \cdot \vpee;
\label{eq:bdops}
\end{eqnarray}
where we have omitted the M1 piece of the $\gamma NN$ vertex, as well
as the portions of the E1 vertex proportional to $\vkay$, the photon
momentum, since they are irrelevant to what follows.  Note that
nucleon recoil is included in the intermediate-state propagators here,
i.e.  they are {\it not} the standard heavy-baryon propagators, which
would give only $1/\omega$ and $-1/\omega$ respectively.  

Evaluation of the operators $\hat{O}^{(a)}$--$\hat{O}^{(c)}$ between
``zero-range'' deuteron wave functions
\begin{equation}
\psi(\vec{r}^{\,})=\sqrt{\frac{\gamma}{2 \pi}} \frac{e^{-\gamma r}}{r},
\label{eq:pionless}
\end{equation}
yields the result~\cite{Ch98}:
\begin{equation}
-\frac{e^2}{M}\left[\frac{4 \gamma}{|\vec{q}^{\,}|}
\arctan\left(\frac{|\vec{q}^{\,}|}{4 \gamma}\right)
+ \frac{2 \gamma^4}{3 M^2 \omega^2}\left\{2 -
\left(1 - \frac{M \omega}{\gamma^2} - i \eta\right)^{3/2}
- \left(1 + \frac{M \omega}{\gamma^2}\right)^{3/2}\right\}\right]
\vepsprime \cdot \veps,
\label{eq:lowenampl}
\end{equation}
with $|{\vec q}^{\,}|=\sqrt{2} \omega (1 - \cos(\theta))$.  (Pieces of
the graphs in Fig.~\ref{fig-bddiags} down by a relative factor of
$\omega/M$ must be omitted in order to obtain (\ref{eq:lowenampl}).)
Taking the limit $\omega \rightarrow 0$ here yields the Thomson limit
for Compton scattering on the deuteron:
\begin{equation}
T_{\gamma d}(\omega=0)=-\frac{e^2}{2 M} \vepsprime \cdot \veps,
\label{eq:deutthom}
\end{equation}
up to corrections due to deuteron binding, which are very small,
being of order $B/(2 M)$.

An important feature of this argument is that in EFT($\not \! \pi$) an
exact accounting of effects due to the scale $\sqrt{M \omega}$ can be
performed.  They produce the second term in Eq.~(\ref{eq:lowenampl}).
As discussed in Ref.~\cite{therestofus}, such contributions are higher
order in our power counting, since if $\omega \sim m_\pi$ they involve
high relative momenta inside deuterium, which means that the deuteron
wave function only enters the integral in regions where it is very
small.  Thus our calculation includes only the first term in
Eq.~(\ref{eq:lowenampl}).  However, as $\omega \rightarrow
\smfrac{m_\pi^2}{M}$ effects from the scale $\sqrt{M \omega}$ become
important, since the deuteron wave function is no longer being
evaluated at a high scale. So, a power counting which regards only
diagram (a) as the leading-order result, as ours does, is not correct
in this very-low-energy regime. Indeed, with diagram (a) alone
included, the amplitude (\ref{eq:deutthom}) is not recovered: a
threshold cross section that is a factor of $M_d/M$ too large results.
This is the source of our difficulty in reproducing lower-energy
deuteron Compton data.

In the absence of a full accounting of the effects of this scale in
$\chi$PT we can only treat more carefully those diagrams which seem to
become enhanced for $\omega \sim \frac{m_\pi^2}{M}$. Of these diagrams the
argument of Ref.~\cite{Ch98} indicates that the most important are
\ref{fig-bddiags}(b) and \ref{fig-bddiags}(c).  Provided we are in the
regime $\omega \gg B$, we can evaluate these diagrams keeping only the
contributions from the pole at $|\vpee^{\,}|^2=M \omega$. Neglecting
the effects of analytic structure at $p \sim \gamma$ and $p \sim
\omega$ , which are suppressed by $B/\omega$ and $\omega/M$
respectively, we get, for \ref{fig-bddiags}(b):
\begin{equation}
T_{s-channel \, \, pole}=
\frac{i \pi e^2}{2 \sqrt{M \omega} M}
\int \frac{d^3p}{(2 \pi)^3} \, \, \psi^*\left (\vpee - \smfrac{1}{2}
\vkayprime \right) \, \, \vepsprime \cdot \vpee
\, \, \delta \left(|{\vec p}\; | - \sqrt{M \omega}\right) \, \,
\veps \cdot \vpee \, \, \psi\left(\vpee - \smfrac{1}{2} \vkay \right),
\label{eq:oq2.5piece}
\end{equation}
If we assume, for the time being, that $\psi$ has only
S-wave components, and again use the hierarchy $\sqrt{\omega M}
\gg \omega = |\vkay^{\,}|=|\vkayprime|$ then:
\begin{equation}
T_{s-channel \, \, pole}=\frac{i e^2}{12 M \pi} \, 
(M \omega)^{3/2} \left|\psi\left(\sqrt{M \omega}\right)\right|^2  \, \vepsprime \cdot \veps.
\label{eq:schann}
\end{equation}
Note that in the strict HB$\chi$PT expansion the imaginary part of the
$\gamma$d amplitude is zero to all orders, provided that $\omega <
m_\pi$. However, Eq.~(\ref{eq:schann}) provides a leading effect in
${\rm Im}(T)$ once we resum the recoil effects in the baryon
propagator. Of course, a number of other mechanisms contribute to
${\rm Im}(T)$, but Eq.~(\ref{eq:schann}) does allow us to estimate
its size.  It indicates that the imaginary part has little effect on
elastic $\gamma$d cross sections at the energies considered in this
work.

An expression similar to Eq.~(\ref{eq:schann}), but with a result that
is ultimately purely real, exists for diagram \ref{fig-bddiags}(c):
\begin{equation}
T_{u-channel \, \, pole}=\frac{e^2}{12 M \pi} \, 
(M \omega)^{3/2}\left|\psi\left(\sqrt{M \omega}\right)\right|^2 
 \, \vepsprime \cdot \veps.
\label{eq:uchann}
\end{equation}
Inserting the Fourier transform of Eq.~(\ref{eq:pionless}) into
Eqs.~(\ref{eq:schann}) and (\ref{eq:uchann}) we recover a result that
agrees with Eq.~(\ref{eq:lowenampl}), as long as $\omega \gg B$. 
More generally, if $\omega \sim \smfrac{m_\pi^2}{M}$, and so $\psi$ is not
small, we see that \ref{fig-bddiags}(b) and (c) are numerically as
important as the Thomson term \ref{fig-bddiags}(a)~\footnote{The role of
similar contributions from on-shell intermediate-state nucleons in
neutral-pion photoproduction on the deuteron has been emphasized by
Wilhelm~\cite{willhelm}.}.

And yet including diagrams \ref{fig-bddiags}(b) and (c) at the same
order as the Thomson term \ref{fig-bddiags}(a) violates the strict
HB$\chi$PT power counting. It amounts to resuming an infinite set of
recoil corrections for the two-nucleon propagator, all of which are
higher-order in the original power counting. Below we refer to this as
the ``very-low-energy resummation''.  To assess the impact of this
resummation which enhances diagrams \ref{fig-bddiags}(b) and (c) in
the region $\omega \sim m_\pi^2/M$ we evaluated these graphs
numerically using the full expression (\ref{eq:oq2.5piece}), together
with its crossed counterpart.  Diagram \ref{fig-bddiags}(c) proves to be
the more important of the two, since it can interfere destructively
with the result for diagram \ref{fig-bddiags}(a), provided that
$|\psi(\sqrt{M \omega})|^2$ does not completely suppress its
contribution.  Ultimately it reduces the cross sections substantially
at 49 MeV, but has little impact on them at 95 (and even 69)
MeV~\cite{therestofus}.  This last result is not surprising, in that at
these higher energies $T_{s-channel \, \, pole}$ is suppressed by
factors of the deuteron wave function at momenta of order $\sqrt{M
m_\pi}$.

It is clearly necessary to do this very-low-energy resummation when
the deuteron is probed at very low photon energies: the baryons can no
longer be treated as static for $\omega \lsim \smfrac{m_\pi^2}{M}$, the
nuclear binding scale, and the Thomson limit (\ref{eq:deutthom}) will
only be recovered if recoil terms for the baryons are included in the
calculation. However, arbitrarily including {\it just}
\ref{fig-bddiags}(b) and (c) ignores the possibility that in this
very-low-energy regime other diagrams may also be enhanced compared to
their $\omega \sim m_\pi$ importance.

In fact, as emphasized in Ref.~\cite{FT81}, getting low-energy
theorems correct can be quite complicated in a theory where the photon
scatters from the deuteron's constituent nucleons and pions, just as
recovering the Thomson limit for the proton can be quite difficult in
a constituent quark model of proton structure~\cite{CK92}.  As pointed
out at the end of Section \ref{sec-pc}, for the EFT discussed here in
the very-low energy region the breakdown into reducible and
irreducible diagrams changes: $K_{\gamma\gamma}$ is no longer
irreducible, and one needs to account for the appearance of the full
two-nucleon propagator $G$ between photon absorption and emission.
Working this out in detail remains an important challenge. Although it
bears pointing out that once $\omega \lsim \smfrac{m_\pi^2}{M}$
EFT($\not \! \pi$) may be a more efficient way to calculate Compton
scattering on deuterium, since at these low energies it is not
necessary to include pions explicitly in the theory in order to
achieve a good description of the data~\cite{griess}.

\subsection{Fits to deuteron Compton data}

\label{sec-gdfits}

Over the past decade three experimental groups have measured elastic
$\gamma$d scattering. The data set includes:
\begin{enumerate}
\item The pioneering
measurements by Lucas at Illinois, who took four data points at 49 MeV
and two at 69 MeV~\cite{lucas}. The four 49 MeV points were taken in
two different runs, and will, in principle, have uncorrelated
systematic errors, therefore we choose to class these four points as
from two different experiments. 

\item Five measurements taken at SAL at photon energies ranging from
84--105 MeV~\cite{SAL}. Data over this energy range was consolidated
into one energy bin, and we quote it at a lab photon energy of 94.5
MeV.

\item The very recent data set from MAX-Lab from
Lund, which includes 9 points at approximately 55 MeV, and another 9
measurements at approximately 66 MeV~\cite{lund}.
\end{enumerate}

\begin{table*}[thbp]
\begin{center}
\begin{tabular}{|l||c|c|c|c|c|}
\hline
  & $E_\gamma$ (MeV) & Angle (deg.) & $\frac{d \sigma}{d \Omega}$ (nb/sr) & Stat. error & Syst. error\\
\hline
\hline
Illinois       &     49.0     &    50.0      &               15.56             &       1.42                &        0.58            \\
            &              &    110.0     &               10.76             &       1.36                &        0.39            \\
\hline
Illinois       &     49.0     &    75.0      &               11.25             &       0.84                &        0.40            \\
            &              &   140.0      &               17.87             &       1.20                &        0.65            \\
\hline
Illinois       &     69.0     &    60.0      &               13.40             &       0.65                &        0.52            \\
           &                &    135.0    &               15.05             &       0.63                &        0.54            \\
\hline
SAL         &     94.5     &    36.8      &             13.3                &       1.3             &        0.9         \\
            &              &    62.7      &         13.5                    &       0.8             &        0.7             \\
            &              &    93.0      &             11.5                &       0.6             &        0.6             \\
            &              &   122.6      &         15.7                &       1.1             &        0.8             \\
            &              &   151.5      &         18.6                &       1.1             &        1.0             \\
\hline
Lund        &     54.9     &    44.3      &     16.6                &       3.3             &        1.8             \\
            &     54.6     &    44.9      &     16.8                &       4.1             &        1.5             \\
            &     55.9     &    50.2      &     13.4                &       2.7             &        1.0             \\
            &     54.6     &    125.0     &     15.7                &       1.5             &        1.3             \\
            &     54.9     &    127.6     &     15.4                &       1.3             &        1.0             \\
            &     55.9     &    131.7     &     15.3                &       2.0             &        1.2             \\
            &     54.9     &    136.3     &     18.4                &       1.7             &        1.6             \\
            &     54.6     &    136.8     &     17.2                &       2.0             &        1.4             \\
            &     55.9     &    137.3     &     21.0                &       3.2             &        2.2             \\
            &     65.6     &    44.4      &     18.6                &       2.4             &        1.4             \\
            &     65.3     &    44.9      &     16.0                &       2.8             &        1.4             \\
            &     67.0     &    50.3      &     15.2                &       1.8             &        1.2             \\
            &     65.3     &    125.3     &     15.3                &       1.3             &        1.4             \\
            &     65.6     &    127.8     &     16.0                &       1.2             &        1.1             \\
            &     67.0     &    131.8     &     14.2                &       1.5             &        1.0             \\
            &     65.6     &    136.5     &     15.1                &       1.7             &        1.3             \\
            &     65.3     &    136.8     &     12.6                &       1.7             &        1.8             \\
            &     67.0     &    137.5     &     15.0                &       2.7             &        1.2             \\
\hline
\end{tabular}
\caption{Deuteron Compton data sets from Illinois~\cite{lucas}, SAL~\cite{SAL},
and Lund~\cite{lund}. Cross sections and angles are quoted in the lab frame
for the Illinois data and the c.m. frame for the SAL and Lund data. Statistical
and systematic errors are reported in nb/sr.}
\label{table-dataset}
\end{center}
\end{table*}

In Table~\ref{table-dataset} we list all of the data, including
systematic and statistical errors. Note that in the table the angles
and cross sections are in the center-of-mass frame for the SAL and
MAX-Lab data, and in the lab frame for the Lucas data. Note also that
since the higher-energy Lund data was taken at almost the same energy
as the 69 MeV Illionis data in what follows we show them on the same
plot, and compare them to calculations at $E_\gamma=67$ MeV. In the
fits by which $\alpha_N$ and $\beta_N$ were determined the energies listed
in Table~\ref{table-dataset} were used to calculate the values for
$(d \sigma/d \Omega)_{\rm theory}$.

\begin{figure}[htb]
\vspace{0.2in}
\epsfxsize=10.0cm
   \centerline{\epsffile{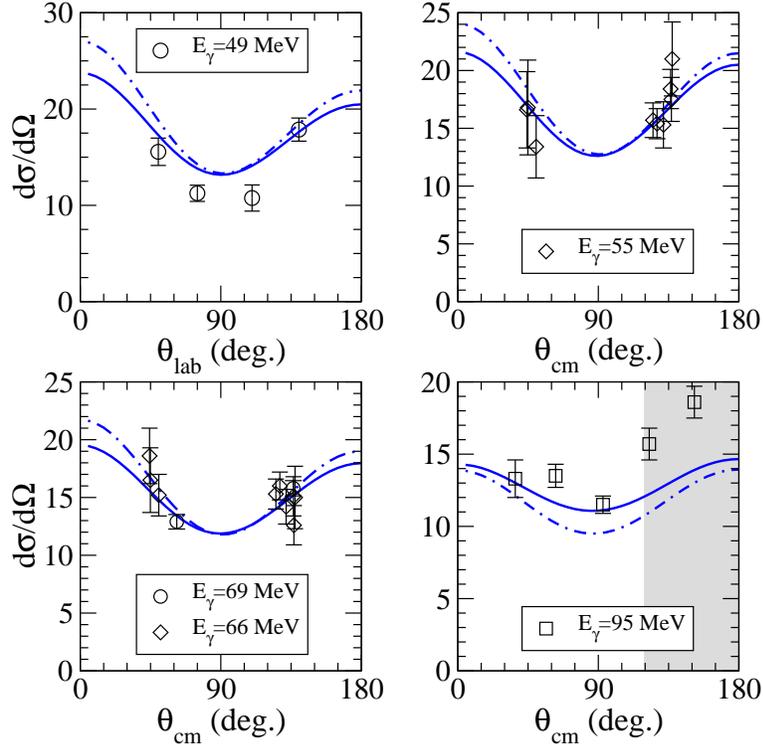}}
\caption {\label{fig-fitIII} Results of the $O(Q^4)$ best fit (including
very-low-energy resummation) to the (lab and c.m. as appropriate)
differential cross sections for Compton scattering on deuterium at
four different lab photon energies: 49, 55, 67, and 94.5 MeV. The
data are from Illinois~\cite{lucas} (circles), Lund~\cite{lund}
(diamonds) and SAL (squares)~\cite{SAL}. The error bars represent the
quoted statistical (only) uncertainties of these measurements. The
solid line is the $O(Q^4)$ calculation with $\alpha_N=13.0 \times
10^{-4}~{\rm fm}^3$, $\beta_N=-1.8 \times 10^{-4}~{\rm fm}^3$ (Fit
III).  The gray area is the region excluded from the fit
($\omega,\sqrt{|t|} > 160$ MeV).  The dot-dashed line is the
(parameter-free) $O(Q^3)$ calculation.}
\end{figure}

These fits seek to minimize the $\chi^2$ defined in
Eq.~(\ref{eq:chisqare}), where systematic errors are accounted for via
a floating normalization. (In the case of the world $\gamma$d data the
index $j$ of Eq.~(\ref{eq:chisqare}) runs from 1 to 5.)  Minimizing
this $\chi^2$ employing the NLO $\chi$PT deuteron wave function of
Ref.~\cite{Ep99} we find a $\chi^2$ per degree of freedom of 2.36,
with a best-fit result of $\alpha_N=13.6 \times 10^{-4}~{\rm fm}^3$
and $\beta_N=0.1 \times 10^{-4}~{\rm fm}^3$. This fit, already shown in
Fig.~\ref{fig-fitI}, has a rather high $\chi^2$. Examining the
cross sections resulting from this calculation shows that the main
contributions to the $\chi^2$ come from the two backward-angle SAL
points, and from the fact that the calculation severely over-predicts
the 49 MeV Illinois data. Both of these problems are associated with
the regime of validity of our calculation, which is valid for $\omega
\sim m_\pi$, but breaks down both at lower energies $\sim m_\pi^2/M$,
where some of the Illinois data were taken, and for values of
$\sqrt{|t|}$ of order the Delta-nucleon mass difference, where the
last two SAL points occur.

In order to deal with this difficulty at higher $\sqrt{|t|}$ we
followed the same procedure as employed above for the proton data and
eliminated points with $\omega,\sqrt{|t|} > 160$ MeV from the
minimization.  This excludes the last two SAL points from the fit,
because our theory apparently does not include all the physics
necessary to describe these points accurately.  Refitting then reduces
the $\chi^2/{\rm d.o.f.}$ to 1.95, with best-fit values of $\alpha_N$
and $\beta_N$ equal to $15.4 \times 10^{-4}~{\rm fm}^3$ and $-2.3
\times 10^{-4}~{\rm fm}^3$ respectively. 

While the restriction on the kinematic range of the fit affects the
central values of $\alpha_N$ and $\beta_N$ it does not reduce the
$\chi^2$ markedly, because the agreement with the 49 MeV data is still
poor.  In contrast, implementing the very-low-energy resummation
(Sec.~\ref{sec-resum}) and then refitting produces a marked
improvement in our description of the data---especially at energies
around 50 MeV.  The $\chi^2/{\rm d.o.f}$ is reduced to 1.33, and the best-fit
values for nucleon electric and magnetic polarizabilities are now $13.0$
and $-1.8$ respectively. This fit (Fit III) is presented in
Fig.~\ref{fig-fitIII}, and is clearly quite good; especially when one
considers that the overall normalization of each set can be adjusted
within their quoted systematic uncertainty.  Omitting the 49 MeV
Illinois data from the fit entirely results in similar central values
for $\alpha_N$ and $\beta_N$, but with a larger statistical
uncertainty.

\begin{figure}[htb]
\vspace{0.2in}
\epsfxsize=10.0cm
   \centerline{\epsffile{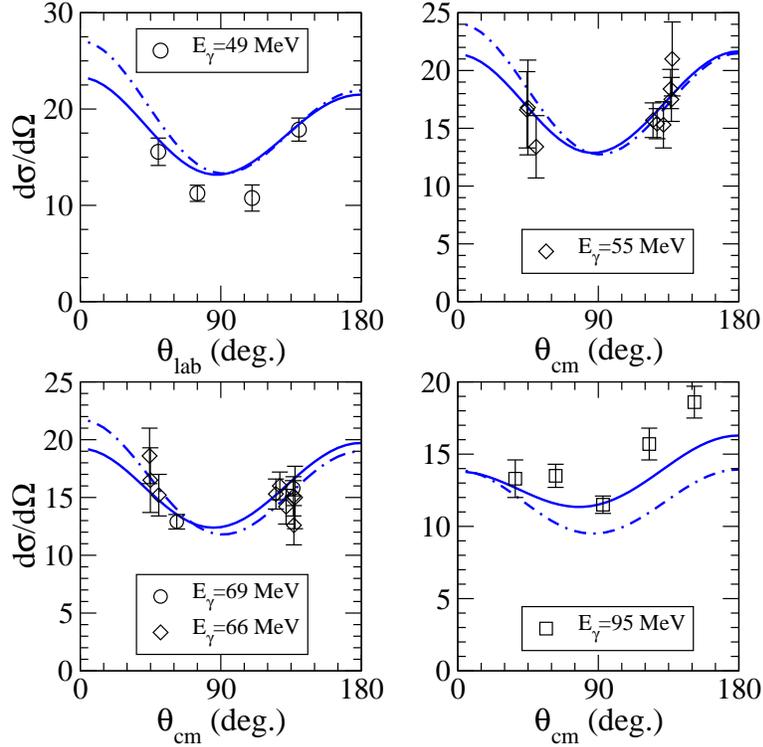}}
\caption {\label{fig-fitIV} Results of the $O(Q^4)$ calculation at
four different lab photon energies: 49, 55, 67, and 94.5 MeV. Data as
in Fig.~\ref{fig-fitI}. The solid line is the $O(Q^4)$ result
(including very-low-energy resummation)
with $\alpha_N=11.5 \times 10^{-4}~{\rm fm}^3$ and $\beta_N=0.3
\times 10^{-4}~{\rm fm}^3$ (Fit IV), while the dot-dashed line is the
(parameter-free) $O(Q^3)$ calculation.}
\end{figure}

Finally, the constraint $\omega,\sqrt{|t|} < 160$ MeV can be modified
to $\omega,\sqrt{|t|} < 200$ MeV, which allows all of the data to be
fitted. The results of that fit are shown in Fig.~\ref{fig-fitIV},
and are presented in Table~\ref{table-results}, 
together with the three other fits discussed so far. 

The numbers differ from those given in our earlier
publication~\cite{allofus} for two reasons. First, there we mistakenly
fitted the SAL data at a c.m. energy of 95 MeV, whereas here we have
done so at a lab energy of 94.5 MeV.  This causes a relatively small
change in the results for most quantities in
Table~\ref{table-results}, but it does increase the central value for
$\alpha_N$ obtained with the Nijm93 wave function by $1.0 \times
10^{-4}~{\rm fm}^3$ over the value quoted in Ref.~\cite{allofus}.  A
larger effect occurs because there was an error in the program used to
generate the $O(Q^4)$ results presented for $\gamma$d scattering in
Ref.~\cite{allofus}. The $\gamma N$ amplitude used there was missing a
factor of $i$ in its spin-dependent parts. This affects the $O(Q^3)$
results by less than 1\%. However, at $O(Q^4)$ the omission of this
factor modifies the interference between the $A_3$--$A_6$ pieces of
the single-nucleon amplitude and the $O(Q^4)$ two-body currents.
Correcting this mistake increases the predicted cross sections by
about 10-15\% with respect to those published in Ref.~\cite{allofus},
and those which appeared in the first version of this
paper~\cite{deepshikha}. We emphasize that this mistake was {\it not}
made in the $\gamma$p calculation reported in Ref.~\cite{allofus} and
discussed in Secs.~\ref{sec-prottheory} and \ref{sec-protres}
above.

\begin{table*}[tbp]
\begin{center}
\begin{tabular}{|c|c|c|c||c|c|c|}
\hline
Fit & Wave & Very-low-energy
& $\omega, \sqrt{|t|} $ &
$\alpha_N$ & $\beta_N$ & $\chi^2$/d.o.f. \\
 & function & resummation? & below & $(10^{-4}~{\rm fm}^3)$ & $(10^{-4}~{\rm fm}^3)$&\\
\hline
\hline
I & NLO $\chi$PT & No  & $200$ MeV & 13.6 & $0.1$ & 2.36 \\
II & NLO $\chi$PT & No  & $160$ MeV & 15.4 & $-2.3$ & 1.95 \\
III & NLO $\chi$PT & Yes & $160$ MeV & 13.0  & $-1.8$ & 1.33\\
IV & NLO $\chi$PT & Yes & $200$ MeV & 11.5  & $0.3$ & 1.69\\
V & Nijm93       & Yes & $200$ MeV & 16.9 & $-2.7$ & 2.87 \\
\hline
\end{tabular}
\caption{Results of our different $O(Q^4)$ EFT extractions of
isoscalar nucleon polarizabilities from low-energy coherent $\gamma$d
scattering data.}
\label{table-results}
\end{center}
\end{table*}

As discussed above, cross sections depend on the choice of deuteron
wave function at the 10-20\% level. This change in the $\gamma$d cross
section has a significant effect on the extraction of $\alpha_N$ and
$\beta_N$. Changing only the wave function yields the difference
between Fit IV and Fit V in Table~\ref{table-results}. The high
$\chi^2$ of Fit V is again due to a failure to reproduce the 49 MeV
data---this time even when the very-low-energy resummation of
Section~\ref{sec-resum} is performed. While the Nijm93 wave function
is not consistent with $\chi$PT, it does have the correct
long-distance behavior. The differences we see are therefore a
consequence of the different short-range behavior of the Nijm93 and
NLO $\chi$PT $NN$ potentials. As such they should be renormalized by
$\gamma NN \rightarrow \gamma NN$ contact operators. However, Weinberg
power counting predicts that the appropriate operators do not appear
until (at least) $O(Q^5)$.  Thus, the degree of variability between
the results with the NLO $\chi$PT and Nijm93 wave functions could be
said to be inconsistent with Weinberg power counting.  Indeed, it may
be necessary to modify the Weinberg power counting so that such
contact operators appear at a lower order than is indicated by naive
dimensional analysis. Further understanding of this issue
is clearly crucial to the use of $\chi$PT as an accurate calculational
tool for low-energy reactions on deuterium, but lies beyond the scope
of this paper. Here we adopt the conservative strategy of assigning
a theoretical error which encompasses both the Nijm93 and NLO
$\chi$PT results.

\begin{figure}[htb]
   \epsfxsize=10.0cm
  \centerline{\epsffile{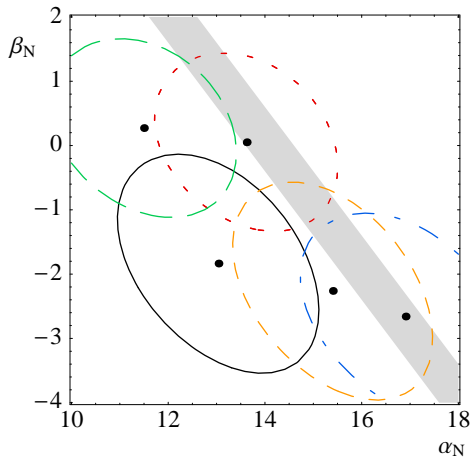}}
\caption{\label{fig-deutellipse} Allowed regions (at the 1-$\sigma$
   level) for isoscalar nucleon electric and magnetic polarizabilities
   (in $10^{-4} \, {\rm fm}^3$) resulting from different $O(Q^4)$ EFT fits
   to low-energy data.  In the order the fits are listed in
   Table~\ref{table-results} the constraints are: fit I (short-dashed,
   red); fit II (medium-dashed, orange); fit III (solid black); fit IV
   (long-dashed, green); fit V (dot-dashed, blue).  For comparison,
   the Baldin sum-rule constraint found by averaging the proton
   and neutron results of Ref.~\cite{Ba98} is shown as the gray shaded
   band.}
\end{figure}

The $\chi^2$ contours for the five fits of Table~\ref{table-results}
are shown in Fig.~\ref{fig-deutellipse}, with the ``best fit'' EFT
(Fit III) giving the 1-sigma region bounded by the solid black line.
Putting these results together we conclude that our central values,
and error bars, for the {\it isoscalar nucleon} polarizabilities 
are:
\begin{eqnarray}
\alpha_N &=& (13.0 \pm 1.9)_{-1.5}^{+3.9} \times 10^{-4} \, {\rm fm}^3,
\nonumber\\
\beta_N &=& (-1.8 \pm 1.9)_{-0.9}^{+2.1} \times 10^{-4} \,
{\rm fm}^3.
\label{eq:Npols}
\end{eqnarray}
The errors inside the brackets are statistical, and those outside
reflect the arbitrariness as to which data are included, and which
deuteron wave function is employed. 

The theoretical error should also include a component from
higher-order ($O(Q^5)$ and above) pieces of the $\gamma NN$ kernel.  A
number of these are taken into account in the sophisticated
potential-model calculations of Refs.~\cite{wilbois,levchuk,jerry}.
Including them in the calculation will alter the central values,
although the theoretical error arising from wave-function dependence
and the variation in the kinematical cuts we impose is large enough
that we expect it encompasses the impact of higher-order terms in the
kernel.

Combining Eqs.~(\ref{eq:protpol1}) and (\ref{eq:Npols}) we see that a
wide range of neutron polarizabilities is consistent with a
model-independent analysis of the current low-energy $\gamma$p and
coherent $\gamma$d data.  However, there is clear statistical evidence
in these data sets for $\alpha_n$ being significantly larger than
$\beta_n$: $\beta_n$ is consistent with zero within our large
error bars. Also, the results presented here provide {\it no} evidence for
significant isovector components in $\alpha$ and $\beta$, in contrast
to some previous results in the literature (see,
e.g. Ref.~\cite{SAL}).

\subsection{Other techniques for measuring $\alpha_n$}

A dispersion sum rule---analogous to the Baldin sum rule for the
proton---that relates the sum of isoscalar nucleon electric and
magnetic polarizabilities to an integral of the deuteron
photo-absorption cross section can be derived, if one assumes that the
single-nucleon contribution dominates the deuteron photoabsorption
cross section. Using this sum rule, and subtracting their result for
$\alpha_p + \beta_p$, the authors of Ref.~\cite{Ba98} found:
\begin{equation}
\alpha_n + \beta_n=(14.40 \pm 0.66) \times 10^{-4} \, {\rm fm}^3.
\label{eq:nucbaldin}
\end{equation}
This result is {\it not} consistent with our ``best fit'' result if
only statistical errors are considered, but if one includes the
theoretical errors in Eq.~(\ref{eq:Npols}) there is no
discrepancy. Moreover, the theoretical error on the sum rule
associated with assuming that the free neutron photoabsorption cross
section can be found by taking the difference of deuteron and proton
data is not clear.

A more sophisticated evaluation of the neutron photoabsorption
integral that uses the neutron multipoles of Ref.~\cite{Ha98} at low energy, 
together with experimental deuteron
and proton photoabsorption cross sections at intermediate energies, and
Regge phenomenology at energies above 1.5 GeV yields~\cite{levchuk}:
\begin{equation}
\alpha_n + \beta_n=(15.2 \pm 0.5) \times 10^{-4} \, {\rm fm}^3.
\label{eq:Baldinn}
\end{equation}
The theoretical error given in Ref.~\cite{levchuk} is arrived at by
employing different neutron pion photoproduction multipoles in the 
sum-rule integrand for the region up to $\omega=500$ MeV.  However, the
theoretical errors from the multipole analyses themselves seem not to
have been assessed. Further, the extent to which the theoretical error
in the high-energy part of the amplitude affects the final result is
not clear to us.

Most direct information on $\alpha_n$ has been obtained by scattering
neutrons on a heavy nucleus and examining the cross section as a
function of energy.  Controversy remains over what result this
technique gives for $\alpha_n$.  In Ref.~\cite{neutpol1} the
value~\footnote{The values measured in these experiments are not the
polarizabilities (often denoted $\bar{\alpha}$) we have been
discussing here.  There is an additional term
$\kappa_n^2/M$~\cite{coon}, with $\kappa_n$ the magnetic moment of the
neutron, in $\bar{\alpha}_n$, which we have added to the experimental
results from neutron-nucleus scattering in order to facilitate direct
comparison. See also Ref.~\cite{Le02}.}
\begin{equation}
\alpha_n=(12.6 \pm 1.5 \pm 2.0)  \times 10^{-4} \, {\rm fm}^3
\label{neutpolexpt2}
\end{equation}
was obtained, which disagrees considerably with the result of the
experiment of Ref.~\cite{neutpol2},
\begin{equation}
\alpha_n=(0.6 \pm 5.0) \times 10^{-4} \, {\rm fm}^3 .
\label{neutpolexpt3}
\end{equation}

Recent data from MAMI~\cite{kossert} exist for quasi-free $\gamma
d\rightarrow \gamma pn$ in the 200 to 400 MeV range. These experiments
update the pioneering work of Ref.~\cite{neutpol3}, which was only
able to set an upper bound on $\alpha_n$ from this process. 
Ref.~\cite{kossert} reports the measurement of
the reaction $\gamma d \rightarrow \gamma p
n$ for laboratory photon scattering angles of 136.2 degrees, and then
uses a theoretical model to extract:
\begin{equation}
\alpha_n - \beta_n=(9.8 \pm 3.6~({\rm stat.}))^{+2.1}_{-1.1}~({\rm syst.})
\pm 2.2 ({\rm model}) \times 10^{-4}~{\rm fm}^3,
\label{neutpolexpt4}
\end{equation}
from their data.  This refines the constraint on $\alpha_n - \beta_n$
obtained somewhat earlier at SAL, where quasi-free Compton deuteron
breakup was measured in the narrower range between 236 and 260 MeV of
photon energy and analyzed using the same theoretical
model~\cite{kolb}. The result (\ref{neutpolexpt4}) is consistent with
the broad range for $\alpha_n - \beta_n$ indicated by our analysis
of the low-energy coherent $\gamma$p and $\gamma$d data.

\section{Conclusion}
\label{sec-conclusion}

Starting from the most general Lagrangian that includes pions and
nucleons and shares the global symmetries of QCD, we have calculated
the amplitude for Compton scattering on the deuteron in \cpt up to
$O(Q^4)$.  This amplitude consists of one- and two-nucleon pieces
averaged over incoming and outgoing deuteron wave functions.  Both the
$\gamma NN$ kernel and the deuteron wave function were obtained from
the same Lagrangian, with a consistent set of parameters.  

The one-nucleon piece of the $\gamma NN$ kernel is the same as the
amplitude for Compton scattering on a nucleon. This was calculated to
$O(Q^4)$ in the usual \cpt power counting in Ref.~\cite{judith}.  That
calculation contains four unknown parameters, which correspond to
short-range contributions to electric and magnetic polarizabilities of
the proton and the neutron.  In this work we fitted the world's
low-energy proton data \cite{pdata,mainz} and determined the two
proton polarizabilities, obtaining values similar to the Particle Data
Group's \cite{PDG}.

We also calculated the two-nucleon kernel to the same order, using the
power counting suggested by Weinberg \cite{weinnp}.  It consists of
various one-pion-exchange contributions with no unknown parameters.
Finally, we used deuteron wave functions fully determined in the same
approach~\cite{Ep99}.  To this order, the deuteron amplitude is
sensitive only to the two isoscalar combinations of the nucleon
electric and magnetic polarizabilities. From the $\gamma$d amplitude
we calculated the differential cross section and compared it with
deuteron data at various photon energies below 100
MeV~\cite{lucas,SAL,lund}.  A good fit was obtained and we extracted
from it results for the two isoscalar polarizabilities, which,
together with our results for proton polarizabilities, allow an
extraction of neutron polarizabilities.  Various theoretical
systematic uncertainties were studied.  

As pointed out in the Introduction,
various sophisticated
calculations of this process exist using potential models
\cite{wilbois,levchuk,jerry}.  Two of the distinctive features of our
work are the use of a consistent $\chi$PT framework throughout, and
the fact that we employ only low-energy data.  We have used {\it only}
data that we believe---on the basis of statistical tests---falls
within the domain of the EFT.  We find that a wide range of neutron
polarizabilities is consistent with a model-independent analysis of
this $\gamma$p and coherent $\gamma$d data. Narrower ranges for the
neutron polarizabilities can be obtained from the data at the expense
of additional assumptions which introduce model dependence.

The potential-model calculations of Refs.~\cite{wilbois,levchuk,jerry}
can be broken up into one- and two-body kernels and wave functions in
the same way ours can.  However, typically the different pieces are
not calculated using a fully consistent field theory.  The wave
function is calculated from a phenomenological potential model that
does not include, {\it e.g.}, the two-pion exchange required by
quantum field theory, but instead employs various single-meson
exchanges. The best calculations strive for consistency of the $\gamma
NN$ kernel with the two-nucleon potential by including in it the same
mesons with the same form factors.  This also leads to a
gauge-invariant $\gamma$d amplitude if the necessary photon seagull
diagrams are included~\cite{levchuk}. Meanwhile the $\gamma N$ piece
of the $\gamma NN$ kernel in Refs.~\cite{wilbois,levchuk,jerry}
includes a polynomical expansion of the $\gamma$N amplitude.  They thus
omit pieces of the $\gamma N$ amplitude which are not analytic in the
photon energy~\cite{therestofus}. On the other hand, the model
calculations include a number of contributions that in our framework
appear only at higher order. There is qualitative, but not
quantiative, agreement between the different calculations and also
between these potential-model calculations and the $\chi$PT
calculation reported on here.

This is perhaps not surprising, since the potential-model calculations
can be understood as various approximations to effective field theory.
Indeed, it is one of the features of EFT that models which do not
badly violate symmetries of QCD should---at low energies---reduce to
the EFT with a particular choice of low-energy constants. (This choice
may, or may not, agree with the choice that is made by QCD.)  Contrary
to what is apparently a common misconception in the literature, there
are no more unknown parameters in our calculation than in these
potential-model calculations.

Various previous extractions of polarizabilities have also differed
from ours in the emphasis given to low-energy data.  Frequently the
Baldin sum rule \cite{baldin}---which relies on certain (reasonable)
assumptions about the behavior of QCD amplitudes at higher energies,
and on model-based interpolation of data--- is used.  Our results are
consistent with the Baldin sum-rule results for $\alpha_p +
\beta_p$~\cite{mainz}, but do not rely on them.  In addition, our
results are consistent, within error bars, with the recent extraction
of $\alpha_N \pm \beta_N$ from the Lund data using the detailed model
of Levchuk and L'vov~\cite{lund,levchuk}. (But see also the values
found using the data of Refs.~\cite{lucas,SAL} and this
model~\cite{SAL,levchuk}.).

There are several ways to improve our present calculation.  First, a
complete treatment of the very-low-energy region would allow inclusion
of low-energy data that could be measured at HI$\gamma$S \cite{HIGS}.
Second, one would like to systematically go to $O(Q^5)$ and higher in
the present energy regime. The significant dependence of our results
for $\alpha_N$ and, to a lesser extent, $\beta_N$, on the choice of
deuteron wave function also warrants further work. It could have
important implications for the formulation of a fully-consistent
nuclear effective theory.

A promising future direction is the addition of an explicit
Delta-isobar field, which would allow us to address higher-energy
data using the same EFT philosophy employed here.  A recent
breakthrough in power counting has facilitated the use of \cpt for
reaction energies up to the Delta peak, in particular in the case
of Compton scattering on the nucleon~\cite{PP03}.  Compton scattering
on the deuteron has very recently been tackled along these lines, and
it seems that the inclusion of an explicit Delta degree
of freedom ameliorates the diffculties we had in reproducing the
backward-angle SAL data~\cite{Hi04}.

Finally, there is, in principle, no obstacle to carrying out EFT
calculations for the quasi-free process $\gamma d \rightarrow \gamma n
p$, and including the existing data \cite{neutpol3,kolb,kossert} in
the extraction of neutron polarizabilities. We hope that these various
improvements will eventually lead to the use of EFT as a
model-independent framework within which nucleon polarizabilities can
be extracted from experimental data gathered in a number of different
reactions.

\bigskip \bigskip

\section*{Acknowledgments}
We thank J.~Feldman, H. Grie\ss hammer, R.~Hildebrandt, and M.~Lucas
for useful discussions, E.~Epelbaum and V.~Stoks for providing us with
deuteron wave functions, and J.~Brower for coding assistance.  We are
also very grateful to D.~Choudhury for pointing out a significant
error in the version of the computer code previously used to generate
results for $\gamma$d scattering. MM, DRP, and UvK thank the Nuclear
Theory Group at the University of Washington and the Institute for
Nuclear Theory program ``Theory of Nuclear Forces and Nuclear
Systems'' for hospitality while part of this work was carried out, and
UvK thanks RIKEN, Brookhaven National Laboratory and the US DOE
[DE-AC02-98CH10886] for providing the facilities essential for the
completion of this work.  This research is supported in part by the US
DOE under grants DE-FG03-97ER41014 (SRB), DE-FG02-93ER40756 and
DE-FG02-02ER41218 (DRP), and DE-FG03-01ER41196 (UvK), by the UK EPSRC
(JM), by Brazil's CNPq (MM), and by an Alfred P. Sloan Fellowship
(UvK).

\appendix
\section{Third-order loop amplitude}
\label{ap-oq3prot}
The third-order one-nucleon loop graphs 
give the following amplitude in a general frame.
The notation $\t_i$ is used for the tensor structures which multiply the
amplitudes $A_i$ of Eq.~(\ref{eq:Ti}); for example $\t_1=\vepsprime\cdot\veps$.
\begin{eqnarray}
T_i&=&{g_A^2 e^2\over 2f_\pi^2}(\t_1+\t_3)J_0[\ombar,m_\pi^2]+\hbox{crossed}\nonumber\\
T_{ii}&=&-{g_A^2 e^2\over 2f_\pi^2}(\t_1+\t_3)\int_0^1\!dx \bigl(J_0[\ombar-x\omega,m_\pi^2]
+J_0[\ombar-x\omega',m_\pi^2]\bigr)+\hbox{crossed}\nonumber\\
T_{iii}&=&{g_A^2 e^2\over 2f_\pi^2}\int_0^1 \!dy \int_0^{1-y}\!dx\Bigl[
(d+1)\t_1J_0[\tilde\omega,m_\pi^2-xyt]\nonumber\\
&&+\Bigl(2(x+y-(d+3)xy)\omega\omega'\t_2
-2V(x,y)\t_1-2(1\!-\!x\!-\!y)\omega\omega'\t_4\Bigr)
J_0'[\tilde\omega,m_\pi^2-xyt]\nonumber\\
&&\qquad+4\omega\omega'xy\Bigl(V(x,y)\t_2+(1\!-\!x\!-\!y)\omega\omega'\t_7
\Bigr) J_0''[\tilde\omega,m_\pi^2-xyt]\nonumber\\
&& \qquad \qquad + 2\Bigl(y i\vec{\sigma}\cdot (\vepsprime \times \qvec) 
\veps \cdot \vkayprime
+x i\vec{\sigma} \cdot (\veps \times \qvec) \vepsprime\cdot\vkay\Bigr)
J_0'[\tilde\omega,m_\pi^2-xyt]\Bigr]+\hbox{crossed}\nonumber\\
T_{iv}&=&-{g_A^2 e^2\over 2f_\pi^2} \t_1\int_0^1 \!dx \Bigl[
(d-1)J_0[(x-\half)\delta\omega,m_\pi^2-x(1-x)t]\nonumber\\&&
\qquad \qquad \qquad -2x(1-x)(t-\delta\omega^2)J_0'[(x-\half)\delta\omega,m_\pi^2-x(1-x)t]\Bigr]
\end{eqnarray}
where $\ombar=(\omega + \omega')/2$,
$\tilde\omega=(\ombar-x\omega-y\omega')$, $\qvec=\vkay-\vkayprime$,
$t=(k'-k)^2=2\omega\omega'(\cos\theta-1)$,
$\delta\omega=\omega-\omega'$.  The integrals $J_0[\omega,m_\pi^2]$,
$J_2[\omega,m_\pi^2]$ and $\Delta_\pi[m_\pi^2]$ have their usual
meanings~\cite{bkmrev}, prime denotes differentiation with respect to
$m_\pi^2$, and \beq V(x,y)=\tilde\omega^2+\omega\omega'-\ombar^2+\half
t(1-x-y+2xy).  \eeq We have also introduced an extra tensor structure:
$\t_7= \vec{\sigma}\cdot(\hat{k'}\times \hat{k})
\vepsprime\cdot\hat{k}\,\veps\cdot \hat{k'}.$ This is not independent;
$\t_7=\sin^2\theta \t_3+\cos\theta \t_5-\t_6$, but as it arises
naturally in the calculations---and enters at too high an order in
$\omega$ to affect the polarizabilities---it is a useful notation.

The notation ``+ crossed" means that to every term is added another with
$\veps\leftrightarrow\vepsprime$, $\vkay\leftrightarrow-\vkayprime$ and 
$\omega \leftrightarrow -\omega'$.  Since the $\t_i$ are all either symmetric
or antisymmetric under 
this transformation, the net effect is to add a term with 
$\omega \leftrightarrow -\omega'$ to the coefficients of $\t_1$ and $\t_2$,
and subtract such a term from the coefficients of $\t_3-\t_7$.

In HB\cpt the difference in photon energies, $\delta \omega$, is of
order $Q^2/M$~\footnote{Strictly speaking, all of the energy arguments
of the $J_0$'s should include $v\cdot p_+$, with $p_+$ the average of
the initial and final heavy-baryon (off-shell) four-momentum.  This
effect, arising from the nucleon kinetic energies, alters the
amplitude only at order $Q^4$, and has been suppressed in the
expressions above. The contribution is included in the fourth-order
amplitudes calculated in  Ref.~\cite{judith} and used here.}. Thus for a
fourth-order calculation only the expansion to first order in
$\delta\omega$ is required.  This gives the following amplitudes
\begin{eqnarray}
T_i&=&{g_A^2 e^2\over 2f_\pi^2}(\t_1+\t_3)J_0[\ombar,m_\pi^2]
+\hbox{crossed}\nonumber\\
T_{ii}&=&-{g_A^2 e^2\over f_\pi^2}(\t_1+\t_3)\int_0^1\!dx J_0[x\ombar,m_\pi^2]
+\hbox{crossed}\nonumber\\
T_{iii}&=&{g_A^2 e^2\over 2f_\pi^2}\int_0^1 \!dy \int_0^{1-y}\!dx\Bigl\{
(d+1)\t_1J_0[\tilde\omega,m_\pi^2-xyt]\nonumber\\&&\qquad\qquad+\Bigl(2(x+y-(d+3)xy)\ombar^2\t_2
-2V(x,y)\t_1\nonumber\\&&\qquad\qquad\qquad\qquad\qquad
-(x+y)\ombar^2(\t_6-\t_5)-2(1\!-\!x\!-\!y)\ombar^2\t_4\Bigr)
J_0'[\tilde\omega,m_\pi^2-xyt]\nonumber\\
&&\qquad \qquad +\half(x-y)^2 \delta\ombar\Bigl(i\vec{\sigma}\cdot
(\vepsprime\times \qvec) \veps\cdot\vkayprime
- i\vec{\sigma}\cdot(\veps\times \qvec) \vepsprime\cdot\vkay\Bigr)
{\partial J_0'[\tilde\omega,m_\pi^2-xyt]\over\partial \tilde\omega}
\nonumber\\&&
\qquad \qquad\qquad+4xyw^2\bigl(V(x,y)\t_2+(1\!-\!x\!-\!y)\ombar^2\t_7\bigr)
J_0''[\tilde\omega,m_\pi^2-xyt]\Bigr\} + \hbox{crossed}
\nonumber\\
T_{iv}&=&-{g_A^2 e^2\over 2f_\pi^2} \t_1\int_0^1 \!dx \Bigl[
(d-1)J_0[0,m_\pi^2-x(1-x)t]\nonumber\\
&&\qquad\qquad\qquad\qquad\qquad\qquad-2x(1-x)\,t\,J_0'[0,m_\pi^2-x(1-x)t]\Bigr]
\end{eqnarray}
where $\tilde\omega=(1\!-\!x\!-\!y)\ombar$, and \beq
V(x,y)=\tilde\omega^2+\half t(1-x-y+2xy).  \eeq There is only one term
proportional to $\delta\omega$ (in $T_{iii}$), and this cancels
against a piece from the fourth-order amplitude (see below).  

\section{Fourth-order loop amplitude}

\label{ap-oq4prot}

The full amplitude in the Breit frame for diagrams 
\ref{protloop}a-\ref{protloop}s can be
obtained from Ref.~\cite{judith}. In a general frame, there are, in
addition to the Breit frame terms, contributions containing $\pbar$,
the average of the incoming and outgoing nucleon momenta.  It is
useful to define the following tensor structures
\begin{eqnarray}
\t_2'&=&\vepsprime\cdot\pbar\; \veps\cdot\vkayprime
+\veps\cdot\pbar\; \vepsprime\cdot\vkay\nonumber\\
\t_4'&=&i\sigbol\cdot(\pbar\times\qvec)\; \vepsprime\cdot\veps\nonumber\\
\t_5'&=&i\sigbol\cdot(\vepsprime\times\vkay) \;\pbar\cdot\veps
-i\sigbol\cdot(\veps\times\vkayprime) \;\pbar\cdot\vepsprime\nonumber\\
\t_6'&=&i\sigbol\cdot(\vepsprime\times\vkayprime)\; \pbar\cdot\veps
-i\sigbol\cdot(\veps\times\vkay) \;\pbar\cdot\vepsprime\nonumber\\
\t_{56}'&=&i\sigbol\cdot(\vepsprime\times\pbar)\; \vkayprime\cdot\veps
-i\sigbol\cdot(\veps\times\pbar) \;\vkay\cdot\vepsprime\nonumber\\
\t_7'&=&i\sigbol\cdot(\hat{k'}\times\hat{k}) \; 
(\vepsprime\cdot\pbar\; \veps\cdot\vkayprime
+\veps\cdot\pbar\; \vepsprime\cdot\vkay) \nonumber\\
\t_7''&=&i\sigbol\cdot(\pbar\times\qvec)\; \vepsprime\cdot\hat{k}\;\veps\cdot\hat{k'}
\label{eq:newts}
\end{eqnarray}
 The notation reflects the structures that these \t's give rise to in
the center-of-mass frame.  Again, not all the structures are
independent, but it greatly simplifies matters to work with this set.

Diagram by diagram, the $\pbar$-dependent contributions are
\begin{eqnarray}
T_a&=&-{g_A^2 e^2\over 2\mn f_\pi^2}(\t_1+\t_3)\,\pq\,
{\partial J_0[\omega,m_\pi^2]\over\partial \omega}+\hbox{crossed}\nonumber\\
T_b&=&{g_A^2 e^2\over 2\mn f_\pi^2}
\Bigl(2(\t_1+\t_3)\,\pq + (\t_2'+ \t_5')\Bigr)\int_0^1\!dx\,
x{\partial J_0[x\omega,m_\pi^2]\over\partial x\omega}+\hbox{crossed}\nonumber\\
T_f&=&-{g_A^2 e^2\over4 \mn f_\pi^2}(1-\tau_3)\,\t_6'\,\omega^{-1}\int_0^1\! dx
J_0[x\omega,m_\pi^2]+\hbox{crossed}\nonumber\\
T_g&=&-{g_A^2 e^2\over4 \mn f_\pi^2}(1+\tau_3)\,\t_6'\,\omega^{-1}\int_0^1\! dx
J_0[x\omega,m_\pi^2]+\hbox{crossed}\nonumber\\
T_h&=&-{g_A^2 e^2\over2 \mn f_\pi^2}\int_0^1 \!dy \int_0^{1-y}\!dx
\left[(1\!-\!x\!-\!y)\pq \left(\t_1(d+3)
{\partial J_0[\tilde\omega,m_\pi^2-xyt]\over\partial\tilde\omega}
\right.\right.\nonumber\\ && \left.\left.
-\Bigl((x+y)\omega^2(\t_6-\t_5)+2(1\!-\!x\!-\!y)\omega^2\t_4+2V(x,y)\t_1
\right.\right.\nonumber\\ && \left.\left.\qquad\qquad\qquad\qquad\qquad\qquad
+2((d+5)x y-x-y)\omega^2\t_2\Bigr)
{\partial J_0'[\tilde\omega,m_\pi^2-xyt]\over\partial\tilde\omega}
\right.\right.\nonumber\\ && \left.\left.
+4xy\omega^2(V(x,y)\t_2+(1\!-\!x\!-\!y)\omega^2\t_7)
{\partial J_0''[\tilde\omega,m_\pi^2-xyt]\over\partial\tilde\omega}\right)
\right.\nonumber\\ && \left.
-\Bigl(\half((d+3)(x+y)-2)\t_2'-\t_4'+\t_6'-\t_5'\Bigr)
{\partial J_0[\tilde\omega,m_\pi^2-xyt]\over\partial\tilde\omega}
\right.\nonumber\\ && \left.
+\Bigl((1\!-\!x\!-\!y)(x+y)\omega^2\t_7'
- 2xy\omega^2\t_7''
+(x+y)V(x,y)\t_2'\Bigr)
{\partial J_0'[\tilde\omega,m_\pi^2-xyt]\over\partial\tilde\omega}\right]
\nonumber\\ && 
-{g_A^2 e^2\over4 f_\pi^2}\delta\omega\int_0^1 \!dy \int_0^{1-y}\!dx(x-y)^2
i\bigl(\vec{\sigma}\cdot\vepsprime\times \qvec\veps\cdot\vkayprime
- \vec{\sigma}\cdot\veps\times \qvec\vepsprime\cdot\vkay\bigr)
{\partial J_0'[\tilde\omega,m_\pi^2-xyt]\over\partial\tilde\omega}
\nonumber\\&&\kern12cm +\hbox{crossed} \nonumber\\
T_i&=&-{g_A^2 e^2\over\mn f_\pi^2}\t_4'\int_0^1\Delta_\pi'[m_\pi^2-x(1-x)t]
\nonumber\\
T_j&=&{g_A^2 e^2\over\mn f_\pi^2}\int_0^1 \!dy \int_0^{1-y}\!dx\Bigl[
2xy\t_7''\omega^2\Delta_\pi''[m_\pi^2-xyt] - \t_4'\Delta_\pi'[m_\pi^2-xyt]
+\hbox{crossed} \nonumber\\
T_k&=&-T_i
\label{eq:genoq4amp}
\end{eqnarray}
Here we have defined $\vec{k}_+$ as the average of the incoming and
outgoing photon momenta.  
The tensors $\t_i'$ have the opposite crossing symmetry to the 
corresponding $\t_i$, as do the products $\pq\,\t_i$.  

Note that the piece in $T_h$ proportional to
$\delta\omega$ comes from $\pbar\cdot\qvec$, and exactly cancels
the piece proportional to $\delta \omega$ in the expansion to fourth
order of the third-order amplitudes.

All the rest of the terms written above can alternatively be generated
by a boost of the third-order Breit-frame amplitude.  The
transformations to do this are as follows:
\begin{eqnarray}
\veps&\to&\veps+ \frac{\veps \cdot \pbar}{M} \, \hat{k};\nonumber\\
\vkay&\to&\vkay-\frac \omega M \, \pbar;\nonumber\\
\omega&\to&\omega-\frac{1}{M} \vec{p}_+ \cdot \vec{k}_+
\label{eq:foboosts}
\end{eqnarray}
with analogous transformations for the other photon.  These boosts are
odd under crossing, which accounts for the change in symmetry
mentioned above.  It is easily seen that the pieces of
(\ref{eq:genoq4amp}) proportional to $\pq$ come from the boost of
$\omega$; they are generated from the fourth-order diagrams which are
just third-order diagrams with an insertion from ${\cal L}^{(1)}$ on
the nucleon line. The other structures come from the boost of the
vectors $\veps$, $\vkay$, etc.  Applying the boost of
Eq.~(\ref{eq:foboosts}) in the third-order amplitude gives the
additional terms needed to ensure equality with the result from
explicit evaluation of the fourth-order diagrams in an arbitrary
frame.

\section{Born terms}

\label{ap-borngen}

In addition to these terms arising from pion loops in a boosted frame,
there are also Born terms.
The complete set of Born terms corresponds exactly to the expansion in
powers of $1/M$ of the result obtained from Dirac nucleons, with the
vertex of an incoming photon of four-momentum $q$ being
\begin{equation}
\Gamma_\mu={\cal Z}\gamma_\mu+{i\kappa\over
2\mn}\sigma_{\mu\nu}q^{\nu},
\end{equation} 
as long as loop renormalizations of bare
parameters such as $\kappa$ are ignored.  

Born terms fall into two categories.  First, there are those which do
not vanish in the Breit frame,
\begin{eqnarray}
T^{\rm Born}_{\rm Breit}
&=&
-{e^2\over 4M^3}\left[
\Bigl(({\cal Z}+\kappa)^2(1+\cos\theta)-{\cal Z}^2\Bigr)(\cos\theta-1)\,\omega^2\,\t_1
-\kappa(2{\cal Z}+\kappa)\cos\theta\,\omega^2\,\t_2\right]\nonumber\\
&& \qquad \qquad \qquad +
{e^2 {\cal Z}^2 \over 2 \mn^2}E_{\rm kin}\,\t_1
\label{eq:TBornBreit}
\end{eqnarray}
where $E_{\rm kin}$ is the average kinetic energy of the nucleons.
The
coefficient of the nucleon kinetic energy term depends on the
normalization of the Dirac spinors in the relativistic case.  As
shown, the normalization is $\bar{u}u=2E/(E+\mn)$, 
which is appropriate for
the standard non-relativistic reduction.
However, the usual covariant treatment of the Compton Born
terms corresponds to the normalization $\bar{u}u=1$, in which
case the last term on the right-hand side of (\ref{eq:TBornBreit})
is not present.

Second, in an arbitrary frame, there are the terms generated by a boost
of the LET pieces of the third-order, Breit-frame amplitude
\begin{eqnarray}
T^{\rm Born}_{\rm boost}&=&-{e^2\over \mn^2 \omega}{\cal Z}^2\t_2'\nonumber\\
&-&{e^2\over 2\mn^3}\Bigl[
\pq\,\Bigl(({\cal Z}+\kappa)^2(\t_3(\cos\theta-1)+\t_4-\t_5)-\kappa^2\t_3
+{\cal Z}({\cal Z}+\kappa)\t_6\Bigr)\nonumber\\&& +\Bigl(\half {\cal Z}^2
- ({\cal Z}+\kappa)^2\Bigr)\t_4'+\kappa^2\t_5'-{\cal Z}({\cal Z}+K)(1-\cos\theta)\t_6'
+\kappa({\cal Z}+\kappa)\t_{56}'\nonumber\\ &&+\smfrac{2}{\omega^2}
{\cal Z}^2\pq\,\t_2'
+2{\cal Z}^2(\cos\theta-1)\veps\cdot\pbar\,\vepsprime\cdot\pbar\Bigr].
\label{eq:borngen}
\end{eqnarray}
The first part of the $\t_4'$ term comes from the Thomson term via a 
Wigner rotation, which is a second-order boost effect; the terms in
the last line are also second-order, with
the second-order boost given by
\begin{eqnarray}
\veps&\to&\veps+\frac {\veps \cdot \pbar} M \, \hat{k}
+\frac {(\pbar \cdot \hat{k})(\veps \cdot \pbar)} {M^2}  \, \hat{k}
-\frac {\veps \cdot \pbar} {2M^2} \, \pbar\nonumber\\
\vkay&\to&\vkay-\frac \omega M \, \pbar
+\frac {\pbar \cdot \vkay} {2M^2} \, \pbar
\label{eq:soboost}
\end{eqnarray}
The angle dependence of these Born terms is more
complicated than that associated with the polarizabilities.

\section{$\gamma N$ kinematics in the $\gamma$d center-of-mass frame}

\label{ap-gammadframe}

In this appendix we consider the evaluation of the single-nucleon
Compton amplitude ``inside'' the deuteron. This necessitates the
evaluation of expectation values:
\begin{equation}
\langle \psi|\t_i'|\psi \rangle,
\label{eq:expval}
\end{equation}
where the $\t_i'$ are the structures defined by Eq. (\ref{eq:newts}),
and $|\psi \rangle$ is a deuteron wave function, calculated for a
deuteron at rest.  In the next appendix we will explain how to correct
such matrix elements for the fact that the initial-state (final-state)
deuteron wave function must be calculated in a reference frame which
is moving with momentum $-\vec{k}$ ($-\vkayprime$).

All $\t_i'$s depend linearly on $\vec{p}_+$, the average of initial-
and final-state nucleon momentum in the frame of choice.  With the
kinematics defined in Fig.~\ref{fig-kinematics} this average, in the
frame where the $\gamma N$ collision takes place, is:
\begin{equation}
\vec{p}_+=\smfrac{1}{2}(\vpee + \vpeeprime) - \smfrac{1}{4}
(\vkay + \vkayprime),
\label{eq:p+}
\end{equation}
with $\vpee^{\, \prime}=\vpee + \smfrac{1}{2} \vec{q}^{\,}$.

The evaluation of the expectation values (\ref{eq:expval}) then
involves integrals of the form:
\begin{equation}
\int \frac{d^3 p}{(2 \pi)^3} \psi^*(\vpee^{\, \prime})
\, \, \vpee \, \, \psi(\vpee^{\,}),
\end{equation}
where all spin labels are suppressed. But, for deuterium, which contains
only $L=0$ and $L=2$ components $\psi(\vpee^{\,})=\psi(-\vpee^{\,})$,
which implies that:
\begin{equation}
\int \frac{d^3 p}{(2 \pi)^3} \psi^*(\vpee^{\, \prime})
\, \, (\vpee + \vpee^{\, \prime})\, \, \psi(\vpee^{\,})=0.
\label{eq:vanish}
\end{equation}
Combining (\ref{eq:vanish}) and (\ref{eq:p+}) we see that the
expectation values (\ref{eq:expval}) may all be evaluated by making
the replacement:
\begin{equation}
\vpee_+ \rightarrow -\smfrac{1}{4}(\vkayprime + \vkay)
\label{eq:replace}
\end{equation}
This substitution is, however, only valid because we are considering
$\gamma$d elastic scattering.

Now, for the third-order $\gamma$d amplitude none of the boosts of the
third-order loops are needed. The only effect arises from boosting the
Thomson term, using the expressions (\ref{eq:foboosts}) for the boosts
of the polarization vectors. (An equivalent result is obtained from
computing the u-channel nucleon pole with E1 $\gamma NN$ vertices.)
This yields the piece of the $\gamma N$ amplitude which appears on the
first line of Eq.~({\ref{eq:borngen}), and so the only $\t_i'$ whose
expectation value is needed at $O(Q^3)$ is $\t_2'$. In the $\gamma$d
center-of-mass frame this contribution to the $O(Q^3)$ $\gamma N$
amplitude can be evaluated using the replacement
(\ref{eq:replace}). The sole effect of such a boost is a 
modification of the function $A_2$:
\begin{equation}
A_2^{\gamma d~{\rm c.m.}}=A_2^{{\rm Breit}} + \frac{\omega}{2 M^2}.
\label{eq:A2gammad}
\end{equation}
Explicit numerical evaluation of the full expectation value of $\t_2'$
shows that the replacement (\ref{eq:replace}) is accurate to a very
good approximation.  Corrections to it arise from terms in
the deuteron wave function suppressed by $\omega^2/M_d^2$. 
Nominally
these are $O(Q^5)$, but in reality they are smaller still,
since they have $M_d^2$ in the denominator.

At $O(Q^4)$ we must also consider expectation values (\ref{eq:expval})
for $i=4,5,6,56,7$, as well as the expectation value of $\t_7''$ and
$\vec{p}_+ \cdot \vec{k}_+$. The pieces of the general-frame amplitude
which arise from boosts of the
third-order amplitude were listed in Appendix~\ref{ap-oq4prot} and on
the second and third lines of Eq.~(\ref{eq:borngen}). They are all
linear in $\vec{p}_+$. Up to corrections of an order higher than we
consider in this work, their contribution to (\ref{eq:expval}) in the
$\gamma$d center-of-mass frame may be obtained via the replacement
(\ref{eq:replace}). This produces the following results for
$A_i^{\gamma d~{\rm c.m.}}$, $i=1 \ldots 6$:
\begin{eqnarray}
A_2^{\gamma d~{\rm c.m.}}&=&A_2^{\rm Breit} + 
\frac{\omega}{2 M}\left[(1 - \cos \theta) A_2^{\rm Breit} 
- A_1^{\rm Breit}\right];\nonumber\\
A_3^{\gamma d~{\rm c.m.}}&=&A_3^{\rm Breit} - 
\frac{\omega}{2 M} \sin^2 \theta (A_4^{\rm Breit} + A_5^{\rm Breit});\nonumber\\
A_4^{\gamma d~{\rm c.m.}}&=&A_4^{\rm Breit} + 
\frac{\omega}{2 M} A_4^{\rm Breit} + \frac{e^2 \omega^2}{8 M^3};\nonumber\\
A_5^{\gamma d~{\rm c.m.}}&=&A_5^{\rm Breit} + \frac{\omega}{2 M}
\left[\left(1 - \frac{1}{2} \cos \theta\right)A_5^{\rm Breit}
- \cos \theta (A_4^{\rm Breit} + A_5^{\rm Breit})\right.\nonumber\\
&& \qquad \qquad \qquad \qquad \qquad \qquad \qquad \qquad \qquad  \qquad \left.
- \frac{1}{2}(A_3^{\rm Breit} - A_6^{\rm Breit})\right];\nonumber\\
A_6^{\gamma d~{\rm c.m.}}&=&A_6^{\rm Breit} + \frac{\omega}{2M}\left[A_4^{\rm Breit} + 
\left(1 - \frac{1}{2} \cos \theta\right) A_6^{\rm Breit} 
+ \frac{3}{2} A_5^{\rm Breit}\right].
\end{eqnarray}

The boost of the Thomson term must be taken care of
separately, since here the second-order boosts of Eq.~(\ref{eq:soboost})
are required. This leads to the fourth-order contributions to the
$\gamma N$ amplitude listed on the last line of
Eq.~(\ref{eq:borngen}). For one of these terms the 
expectation value
\begin{equation}
\langle \psi|\veps \cdot \vec{p}_+ \, \, \vepsprime \cdot \vec{p}_+|\psi \rangle
\end{equation}
enters. It cannot be calculated by symmetry arguments, but must be
retained and computed by explicit numerical integration.  These
$O(Q^4)$ terms are included in our calculation of the $\gamma$d
amplitude, although since they are suppressed by $1/M^2$ relative to
leading their contribution is numerically small.

Lastly, in the general-frame, $O(Q^4)$, Compton amplitude discussed in
Appendices~\ref{ap-oq4prot} and \ref{ap-borngen} there are a number of
terms proportional to $\vec{p}_+ \cdot \vec{k}_+$.  These terms all
arise from applying the first-order boost (\ref{eq:foboosts}) to the
Breit-frame photon energy $\omega_b$.  Using the replacement
(\ref{eq:replace})---which is valid for the terms linear in
$\vec{p}_+$---we may replace $\vec{p}_+ \cdot \vec{k}_+$ by
$-|\vec{k}_+|^2/2$.  Consequently all but one of the these terms in
the fourth-order amplitude may be included in the final result
simply by employing the energy
\begin{equation}
\omega_{{\rm half}b} \equiv \omega_b - \frac{1}{4M} \omega^2(1 + \cos \theta),
\end{equation}
when evaluating the third-order $\gamma N$ amplitude. This is the
strategy we adopted in our calculation.

The only term in the $O(Q^4)$, general-frame, $\gamma N$ amplitude for
which the use of $\omega_{\rm halfb}$ in the $O(Q^3)$ amplitude does
{\it not} yield the correct $\vec{p}_+ \cdot \vec{k}_+$ pieces of the
$O(Q^4)$ $\gamma$d amplitude is the penultimate term in Eq.~(\ref{eq:borngen}):
\begin{equation}
-\frac{e^2 {\cal Z}^2}{M^3 \omega^2} \vec{p}_+ \cdot \vec{k}_+ \t_2',
\end{equation}
which is second-order in $\vec{p}_+$.
The error that results from evaluating this piece of $T_{\rm
boost}^{\rm Born}$ using $\omega_{{\rm half}b}$ in the third-order
amplitude is of relative order $1/M^2$. Ultimately, we expect a number
of $O(Q^5)$ effects to be more important than this particular
difference.

\section{Boosting the deuteron wave function}

At $O(Q^4)$ we also have to take into account that the deuteron is
not in its rest frame in either the initial or final state.

Consider the two-body matrix element of the operator $\hat{O}$:
\begin{equation}
\langle \hat{O} \rangle_{cm}=
\langle \vpeeprime, \, -\vkayprime|\hat{O}|\vec{p}, \, -\vec{k}\rangle,
\end{equation}
where we are employing a basis of two-nucleon states expressed
in terms of relative and center-of-mass momenta, and we have chosen
to work on the $\gamma$d center-of-mass frame, where 
the initial (final) momentum of the two-nucleon system is $-\vec{k}$
($-\vkayprime$). Using the free boost operator $\hat{\chi}_0$ 
and working to first order in the boost (which is all that
is necessary at this order) we write:
\begin{equation}
\langle \hat{O} \rangle_{cm}=\langle \hat{O} \rangle_{rest}
+ \langle \vec{p}^{\, \prime}, \, \vec{0}|i[\hat{\chi}_0(-\vec{k}') \hat{O}
- \hat{O} \hat{\chi}_0 (-\vec{k})]|\vec{p}, \, \vec{0}\rangle.
\end{equation}
where:
\begin{equation}
\langle \hat{O} \rangle_{rest}=\langle \vec{p}^{\, \prime}, \, \vec{0}|
\hat{O}|\vec{p}, \, \vec{0} \rangle
\end{equation}
is the zeroth-order result.

The free boost operator $\hat{\chi}_0(\vec{P})$ is  given by~\cite{AA97}:
\begin{equation}
\hat{\chi}(\vec{P})=\frac{1}{8 M^2} \left\{ -\frac{1}{2} \left[\vec{r} \cdot
\vec{P} \, \, \vec{p} \cdot \vec{P} + \vec{p} \cdot \vec{P} \, \, \vec{r} \cdot
\vec{P}\right] + (\vec{\sigma}_1 - \vec{\sigma}_2) \times \vec{p}
\cdot \vec{P}\right\},
\end{equation}
where $\vec{r}$ and $\vec{p}$ are to be interpreted as quantum-mechanical
operators, i.e.  they do not commute with each other.
So we see that we can write
\begin{equation}
\langle \hat{O} \rangle_{cm}=\langle \hat{O} \rangle_{rest}
+ \langle \hat{O} \rangle_{\hat{\chi}_r} + \langle \hat{O} \rangle_{\hat{\chi} \sigma},
\end{equation}
where, upon evaluation:
\begin{equation}
\langle \hat{O} \rangle_{\hat{\chi}_r}
=\frac{1}{8M^2} \left(\omega^2 + \vec{p}^{\, \prime} \cdot \vkayprime
\, \, \nabla_{p'} \cdot \vkayprime + \vec{p} \cdot \vec{k} \, \, 
\nabla_{p} \cdot \vec{k}\right) \langle \hat{O} \rangle_{rest},
\end{equation}
and, if $\hat{O}$ is spin-independent:
\begin{equation}
\langle \hat{O} \rangle_{\hat{\chi}_\sigma}=
\frac{i}{8 M^2} \left[
\langle S=1|(\vec{\sigma}_1 - \vec{\sigma}_2) \cdot
(\vkayprime \times \vpeeprime - \vkay \times \vpee^{\,})
|S=1\rangle\right]
\langle \hat{O} \rangle_{rest},
\end{equation}
which vanishes because the $|S=1\rangle$ wave function is symmetric
under the interchange of the spins of particles one and two. 

Since both $\langle \hat{O} \rangle_{\hat{\chi}_r}$ and $\langle
\hat{O} \rangle_{\hat{\chi}_\sigma}$ are suppressed by $1/M^2$ at
$O(Q^4)$ we need only consider what impact they have on the evaluation
of the $O(Q^2)$ $\gamma NN$ kernel.  Thus, we now move to the specific
case where the operator $\hat{O}$ is the $O(Q^2)$ $\gamma N$
amplitude:
\begin{equation}
\langle \hat{O} \rangle_{rest}=-\frac{e^2}{M} \delta^{(3)}
\left(p' - p - \smfrac{1}{2}(k - k')\right).
\end{equation}
After some algebra we find that:
\begin{equation}
\langle \hat{O} \rangle_{rest} + \langle \hat{O} \rangle_{\hat{\chi}_r}
=-\frac{e^2}{M} \left(1 + \frac{\omega^2}{8M^2}\right)
\delta^{(3)}\left(p' - p - \smfrac{1}{2}(k_{eff} - k'_{eff})\right),
\end{equation}
where:
\begin{equation}
\vec{k}_{eff}=\vec{k} + \frac{\vec{k} \cdot (\vec{p} + \vec{p}^{\, \prime})}{8 M^2}
\vec{k} - \frac{\vec{k} \cdot \vec{k'}}{16 M^2} \vec{k},
\end{equation}
with a similar result for $\vec{k}'_{eff}$. Using the fact that 
$\vec{p}^{\, \prime}$ is constrained to be $\vpee + \smfrac{1}{2}\vec{q}$
when we are evaluating this particular operator we find:
\begin{equation}
\vkayprime_{eff} - \vec{k}_{eff}=\vkayprime\left(1 - \frac{\omega^2}{8 M^2}
(1 - \cos\theta)\right) - \vec{k} + \frac{\vkayprime\cdot \vec{p}}{4 M^2} \
\vkayprime - \frac{\vec{k} \cdot \vec{p}}{4 M^2} \vec{k}.
\label{eq:contracted}
\end{equation}

If we write $\vec{p}$ as $-\smfrac{1}{4} \vec{q}$ plus terms
which vanish upon taking the expectation value at the order we work to
here, Eq.~(\ref{eq:contracted}) becomes:
\begin{equation}
\vec{k}^{\, '}_{eff} - \vec{k}_{eff}=\vec{q} \left(1 - \frac{|\vec{q}^{\,}|^2}{16
M^2}\right),
\end{equation}
which coincides with Adam and Arenh\"ovel's result for the one-body charge
operator in the Breit frame~\cite{AA97}. This effective reduction in
$\vkayprime - \vec{k}$ can be interpreted
as a length-contraction effect. However, at the energies
considered here it is very small: at $\theta=180$ degrees,
$\omega=95$ MeV, it produces only a 0.3\% change in $|\vec{q}^{\,}|$.

\end{document}